# CHARACTERIZING DEPOSITS EMPLACED BY CRYOVOLCANIC PLUMES ON EUROPA

**Dr. Lynnae C. Quick**[*]

NASA Goddard Space Flight Center

Planetary Geology, Geophysics and Geochemistry Laboratory

8800 Greenbelt Road

Greenbelt, MD 20771

[*]Corresponding Author at: Lynnae.C.Quick@nasa.gov

**Dr. Matthew M. Hedman**

University of Idaho

Department of Physics

875 Perimeter Drive, MS 0903

Moscow, ID 83844

**Abstract:** In the absence of direct observations of Europa's particle plumes, deposits left behind during eruptive events would provide the best evidence for recent geological activity, and would serve as indicators of the best places to search for ongoing activity on the icy moon. Here, we model the morphological and spectral signatures of europan plume deposits, utilizing constraints from recent Hubble Space Telescope observations as model inputs. We consider deposits emplaced by plumes that are 1 km to 300 km tall, and find that in the time between the Galileo Mission and the arrival of the Europa Clipper spacecraft, plumes that are < 7 km tall are most likely to emplace deposits that could be detected by spacecraft cameras. Deposits emplaced by larger plumes could be detected by cameras operating at visible wavelengths provided that their average particle size is sufficiently large, their porosity is high, and/or they are salt-rich. Conversely, deposits emplaced by large plumes could be easily detected by near-IR imagers regardless of porosity, or individual particle size or composition. If low-albedo deposits flanking lineated features on Europa are indeed cryoclastic mantlings, they were likely emplaced by plumes that were less than 4 km tall, and deposition could be ongoing today. Comparisons of the sizes and albedos of these deposits between the Galileo and Europa Clipper missions could shed light on the size and frequency of cryovolcanic eruptions on Europa.

## 1. Introduction

The youthful and heavily fractured surface of Jupiter's moon Europa indicates that it has been geologically active in the relatively recent past. Multiple Hubble Space Telescope (HST) observations of large, water-vapor-dominated plumes suggest that the icy moon may currently be geologically active, with water vapor being actively vented into space [*Roth et al.*, 2014; *Sparks et al.*, 2016; 2017]. Additionally, a reanalysis of Galileo data suggests that several instruments aboard the spacecraft may have detected plume activity during low-altitude flybys of the moon [*Jia et al.*, 2018]. Meanwhile, *Fagents et al.* [2000] suggested that low-albedo deposits lying along lineaments and surrounding lenticulae on the icy moon could be ballistically-emplaced mantlings of cryoclastic material. Assuming that eruptive events were driven by volatiles such as CO, $CO_2$, $SO_2$ and $NH_3$, they constrained the dimensions of plumes that could have emplaced these deposits. Focusing solely on the particle component of Europan plumes, *Quick et al.* [2013] extended this analysis by constraining likely optical depth values and eruption lifetimes for these plumes. These authors assumed that the putative deposits imaged by the Galileo spacecraft were on the order of 1-10 m thick and consisted solely of particles with 0.5 $\mu$m radii. Similar to the results of *Fagents et al.* [2000], they concluded that the dimensions of the dark deposits on Europa were consistent with emplacement by plumes that were 2.5 - 26 km tall. Additionally, *Quick et al.* [2013] found that plumes with optical depths $\geq$ 0.04 were most likely to be detected by spacecraft operating at visible wavelengths. This optical depth value corresponded to *I/F* $\geq$ 0.07, where *I/F* is a standardized measure of the plume's reflectance, and particle column densities of 1.88 x 10$^{-6}$ kg/m$^2$ [*Quick et al.*, 2013].

Although recent HST observations suggest the presence of large vapor plumes [*Roth et al.*, 2014], previous searches for plumes in the Galileo dataset yielded null results, suggesting that venting on Europa is not dominated by sizeable eruptions, and/or that eruptions may be sporadic in nature [*Phillips et al.*, 2000]. Hence if large plumes are outliers, and most plumes on Europa are indeed small in stature as has been suggested by image analysis [*Phillips et al.*, 2000; *Quick et al.*, 2010; *Bramson et al.*, 2011] and modeling [*Fagents et al.*, 2000; *Quick et al.*, 2013], it is possible that active venting would have been missed by spacecraft cameras. Conversely, if plumes erupted at a time when Galileo was not observing Europa's limb, or, if large plumes on Europa are similar to Io's proposed "stealth plumes" [*Johnson et al.*, 1995], i.e., primarily composed of vapor-phase volatiles, then cameras would have missed eruptions altogether. Moreover, the extent to which thermal observations can be relied upon to reveal plume activity on Europa is unclear [*Rathbun and Spencer*, 2018; 2019]. Thus, in the absence of direct, in-situ observations of active plumes, the identification of fresh plume deposits may be the only tangible evidence of recent and/or ongoing activity on Europa.

Searching for evidence of recent activity on Europa and determining the extent of material exchange between the ice shell and ocean are key subgoals of NASA's Europa Clipper Mission [*Europa Science Definition Team*, 2012; *Phillips and Pappalardo*, 2014; *Pappalardo et al.*, 2015; *Turtle et al.*, 2016; 2019]. Given the large uncertainties surrounding the scale and frequency of eruptions on Europa, constraining the dimensions and deposition rates of plume deposits may be the best way to quantify current activity. Nevertheless, quantitative examinations of the appearance of these deposits, and the potential for their detection by instruments on Europa Clipper have been limited (but see *Southworth et al.* [2015]). While the presence of plume deposits may be indicative of the presence of liquid water at shallow levels in the ice shell, plume deposit dimensions could help to constrain rates of material exchange between the surface and subsurface and provide a baseline from which the properties of the plumes that emplaced them could be extrapolated. In addition, plume particle size distributions and deposition rates could be utilized to predict the character of localized albedo changes caused by the emplacement of cryoclastic particles on Europa's surface. Moreover, as is the case on Io where localized venting and the subsequent deposition of pyroclastic deposits occurs contemporaneously with effusive eruptions [*Geissler et al.*, 2004], cryoclastic particle deposition on Europa could occur within the same timespan that cryolava is extruded onto the surface. Further, owing to their minimum exposure to Europa's severe radiation environment, fresh plume deposits may contain and preserve organic compounds (e.g., see *Nordheim et al.*, [2018]). The characterization of candidate plume deposits is therefore a crucial step in gauging both the rates of occurrence, and probable locations of, cryovolcanic activity on Europa. Their potential to contain biomarkers also makes their identification a critical step in constraining Europa's habitability and astrobiological potential.

Here we employ the properties of candidate Europan plumes, gathered from observational data and modeling, to characterize the dimensions, deposition rates, particle size distributions, and spectral properties of their resulting deposits on Europa's surface. Europa's plumes may be generated by exsolution of $CO_2$, $SO_2$, etc. in fractures propagating from the ocean to the surface [*Crawford and Stevenson*, 1988; *Fagents et al.*, 2000], or similar to Enceladus' plumes, by $H_2O$ boiling at the surface of a water column exposed to the vacuum of space [cf. *Berg et al.* 2016]. It is unclear whether stress states in Europa's crust would allow fractures to extend from the surface directly to the ocean, especially if the icy crust is tens of km thick, or if the ice overlying the ocean is sufficiently ductile [*Crawford and Stevenson*, 1988; *Gaidos and Nimmo*, 2000; *Fagents*, 2003]. Hence we assume the latter case, in which plumes are generated by the boiling of a water column at 273 K. In this case, fractures that expose fluids to Europa's zero-pressure surface environment may be connected to fluid reservoirs that exist at shallow levels in the crust [c.f. *Gaidos and Nimmo*, 2000; *Fagents*, 2003; *Schmidt et al.*, 2011].

In Section 2, we introduce the dynamical and spectral models that were used to perform our analyses. We present our dynamical and spectral modeling results in Section 3 and discuss the implications of these results in Section 4. Finally, we conclude in Section 5 by placing constraints on the eruption rates that are necessary to emplace candidate plume deposits along lineaments on Europa. We also summarize the specifications of visible and near-IR imagers on the Europa Clipper spacecraft and comment on their ability to detect plume deposits on the icy moon. For reference, all variables that are utilized in our dynamical and spectral models are listed in Table 1.

**Table 1**. Plume and Deposit Parameters

| Symbol | Parameter | Value | Unit |
|---|---|---|---|
| $A$ | Deposit area | __ | km² |
| $A_1$ | Brightness coefficient | __ | |
| $c_p$ | Specific Heat at Constant Pressure | __ | J/K-kg |
| $c_v$ | Specific Heat at Constant Volume | __ | J/K-kg |
| $D$ | Deposition rate | | m/s |
| $g$ | Acceleration due to gravity on Europa | 1.31 | m/s² |
| $H$ | Maximum plume height | 1 - 300 | km |
| $H_P$ | Maximum height of individual plume particles | __ | km |
| $I/F$ | Plume Brightness | __ | __ |
| $I/V$ | Ice to vapor mass ratio | | |
| $k_B$ | Boltzmann's Constant | 1.28 x 10⁻²³ | J/K |
| $L$ | Particle collision length | 0.1 | m |
| $m$ | Mass of a water molecule | 2.99 x 10⁻²⁶ | kg |
| $m_w$ | Molar mass of a water molecule | 1.8 x10⁻² | kg/mol |
| $m_p$ | Mass of individual plume particles as a function of size | | kg |
| $M$ | Total mass of plume particles | __ | kg |
| $M_P$ | Total mass of particles of a particular size | | |
| $M_v$ | Total mass of water vapor in the plumes | __ | kg |
| $n$ | Real index of refraction | __ | __ |
| $n_f$ | Mass fraction of driving volatile | __ | __ |
| $n_{gas}$ | Weight percent of driving volatile | 0.1-100 | % |
| $N$ | Total number of particles in the plume | __ | __ |
| $N_A$ | Avogadro's Number | 6.022 x 10²³ | kg/mol |
| $N_p$ | Number of particles of a certain size in the plume | __ | __ |
| $N_w$ | Total number of water molecules in the plume | __ | __ |
| $r_c$ | Critical radius of plume particles | __ | μm |
| $R_g$ | Ideal gas constant | 8.314 | J/mol-K |
| $R_p$ | Maximum range of plume particles | | km |
| $r_p$ | Plume particle radius | 0.5 - 3 | μm |
| $r_{eff}$ | Effective average particle size in deposit | | μm |
| $S$ | Effective average scattering length in regolith | | μm |
| $t_{deposit}$ | Deposit accumulation time | __ | hour; day; year |
| $t_R$ | Particle residence time | | s |
| $T$ | Eruption temperature | 240 | K |
| $T_D$ | Deposit thickness | __ | m |
| $T_{total}$ | Total deposit thickness per eruption | | mm |
| $v = v_{gas}$ | Maximum plume eruption velocity | | m/s |
| $v_p$ | Plume particle velocity | __ | m/s |
| $V_p$ | Volume of plume particles | __ | m³ |
| $\alpha$ | Absorption Coefficient | | μm⁻¹ |
| $\beta$ | Condensation coefficient | 0.2 | __ |
| $\gamma$ | Ratio of specific heats of water vapor ($c_p/c_v$) | 1.334 | __ |
| $\theta$ | Particle eruption angle | 45 | ° |
| $\kappa$ | Molecular weight of water vapor | 1.8 x 10⁻² | kg/mol |
| $\kappa_r$ | Imaginary index of refraction | | |
| $\lambda$ | Wavelength of light | 1-2.5 | μm |
| $\rho_{gas}$ | Density of water vapor | 4.85 x 10⁻³ | kg/m³ |
| $\rho_p$ | Density of plume particles | 920 | kg/m³ |
| $\phi$ | Deposit porosity | 0.5; 0.9 | __ |

## 2. Materials and Methods

**2.1 Dynamical Model**

We utilize an analytical model, based on the work of *Quick et al.* [2013], to estimate the dimensions of deposits generated by a variety of plumes. While our model represents a simplification of the dynamics associated with plume particle deposition on Europa, such a simplified model allows us to explore the full parameter space as it relates to deposit dimensions and the sizes of plumes that may have emplaced them. As in *Quick et al.* [2013], we have assumed that eruptions are steady, i.e., the number of particles supplied to, and the particle discharge rate from, any plume as a function of time are both constant [*Parfitt and Wilson*, 2008]. We calculate the maximum travel distance of each plume particle assuming that particles are launched from eruptive sources regions at an angle of 45° from the horizontal and travel along ballistic trajectories. As suggested by HST observations, we assume that water vapor is the main volatile that drives eruptions [*Roth et al.*, 2014; *Sparks et al.*, 2016; 2017], and have utilized the methods of *Fagents et al.* [2000] and *Quick et al.* [2013] to constrain general plume particle eruption dynamics, assuming that plumes on Europa follow ballistic trajectories after ejection from the vent, a scenario which has been previously modeled for plumes on Enceladus [*Degruyter and Manga*, 2011].

We assume that plumes and their deposits consist of particles that range from 0.5 μm to 3μm in radius, consistent with the size range of plume particles on Enceladus and Io [*Cook et al.*, 1981; *Strom et al.*, 1981; *Collins et al.*, 1981; *Porco et al.*, 2006; *Spahn et al.*, 2006; *Kempf et al.*, 2008; *Postberg et al.*, 2008; *Hedman et al.*, 2009; *Kieffer et al.*, 2009; *Ingersoll and Ewald*, 2011]. Although plume particles may contain salts and other compounds (c.f. [*Postberg et al.*, 2008; 2011; 2018; *Hsu et al.*, 2015; *Porco et al.*, 2017]), for the sake of simplicity we assume that all particles are solely composed of water ice.

*2.1.1 Plume Parameters*

As eruptions are assumed to be vapor driven, the maximum velocity of the gas-particulate mixture upon eruption may be expressed as:

$$v = v_{gas} = \sqrt{\frac{2n_f R_g T \gamma}{\kappa(\gamma-1)}} \qquad (1)$$

[*Wilson and Head*, 1983; *Fagents et al.*, 2000] where $n_f$ is the mass fraction of gas in the erupting plume, $R_g$ = 8.314 J/mol-K is the universal gas constant, and $T$ is the gas temperature at the time that the erupted material expands into Europa's zero-pressure surface environment. $\gamma = c_p/c_v$ = 1.334 and $\kappa$ = 1.8 x $10^{-2}$ kg/mol represent the ratio of specific heats and the molecular weight, respectively, of the volatile driving the eruption, which in this case is water vapor.

Equation (1) has been utilized in previous work to describe the velocities reached by grains during explosive eruptions on both the moon and Europa [*Wilson and Head*, 1983; *Fagents et al.*, 2000]. In those cases, eruptions included volatile contents as low as 0.07 wt%, and 0.09 wt% ($n_f$ = 7 x $10^{-4}$ and 9 x $10^{-4}$), respectively, and plumes were 0.4 km to 1 km tall [*Fagents et al.*, 2000; L. Wilson, personal communication]. Owing to their low volatile contents, these plumes would have had very high solid to vapor mass ratios. In such particle-rich plumes, momentum is transferred from the gas to the particles in order to keep the latter in motion. Particles will remain interspersed in a dense column of gas near the vent, enabling efficient coupling between the gas and grains, especially in cases when the solids to gas ratio is at least on the order of 10 [*Yeoh et al.*, 2015; *Mahieux et al.*, 2019]. Icy particles in these plumes would therefore be in constant contact with the driving gas [*Yeoh et al.*, 2015; *Berg et al.*, 2016]. Equation (1) describes the velocity of all of the icy particles upon eruption as long as they remain coupled with the driving volatile (L. Wilson, personal communication). Tables 2 and 3 illustrate that plumes ≤ 25 km tall will have low water vapor content and will therefore have high enough ice to vapor mass ratio values ($I/V$) to be within this limit (see Section 3). Thus, the velocity of icy particles, regardless of particle size, in plumes ≤ 25 km tall, is adequately described by (1).

Conversely, large plumes ($H \geq 50$ km) will have high vapor contents and relatively low *I/V* (Tables 2 & 3). Individual particles in plumes with *I/V* ≤ 1 will interact more with the walls of the fissure than with other particles. As a consequence, their upward motion will be dependent upon how often they collide with the fissure walls before they can be reaccelerated by the driving gas [*Schmidt et al.*, 2008; *Yeoh et al.*, 2015]. Hence, while (1) is adequate to describe the motion of icy particles in small, particle-packed plumes, we must consider the velocity of particles as a function of size for large plumes with low *I/V*. The dynamics of icy particles in large Europan plumes may be similar to the dynamics of particles in Enceladus' plumes. We therefore apply the dynamical model of *Schmidt et al.* [2008] to obtain velocity distributions, as a function of particle size, for particles in plumes that are ≥ 50 km tall.

*Schmidt et al.* [2008] illustrated that plume particle speeds on Enceladus are affected by wall collisions and that particle acceleration is dependent upon gas density and particle size. Likewise, ionian plume dynamics are also dependent upon particle interactions with the driving gas [*Zhang et al.*, 2004; *Geissler and Goldstein*, 2006]. The average velocity $\langle v_p \rangle$ of plume particles as a function of particle radius, $r_p$, in large, low *I/V* plumes is:

$$\langle v_p(r_p) \rangle = v \bigg/ \left(1 + \frac{r_p}{2r_c}\right) \tag{2}$$

[*Schmidt et al.*, 2008], where $r_c$, the critical radius, is expressed as:

$$r_c = \frac{\rho_{gas}}{\rho_p} \sqrt{\frac{8 k_B T}{\pi m}} \left[1 + \frac{\pi}{8}(1-\beta)\right] \frac{L}{v} \tag{3}$$

[*Schmidt et al.*, 2008]. Here $\rho_{gas}$ and $\rho_p$ are the density of water vapor and icy particles, respectively, $k_B$ = 1.28 x 10$^{-23}$ J/K is Boltzmann's constant, and $m$ = 2.99 x 10$^{-26}$ kg is the mass of one water molecule. $\beta$ is a condensation coefficient, a quantity that represents the adsorption of water molecules by ice grains [*Shaw et al.*, 1999; *Batista et al.*, 2005]. $L$ is the collision length, which represents the characteristic distance that particles are able to travel between collisions with the walls of fractures that transport plume material to the surface. According to *Schmidt et al.* [2008] particles with $r_p < r_c$ travel with $<v_p> \sim v$, while particles with $r_p > r_c$ have wide velocity distributions and their maximum velocity peaks at a speed $v_{max} < v$. We assume that all particles are spherical and are solely composed of water ice [*Degruyter and Manga*, 2011; *Quick et al.*, 2013; *Hedman et al.*, 2018], so that $\rho_p = 920$ kg/m$^3$. As in *Schmidt et al.* [2008] we have assumed that $\rho_{gas} = 4.85 \times 10^{-3}$ kg/m$^3$ is the density of water vapor.

Assuming ballistic trajectories for plume particles [*Fagents et al.*, 2000; *Quick et al.*, 2013], the maximum height, $H_p$, that individual plume particles reach above the surface, regardless of whether their motion is best described by (1) or (2), is:

$$H_p = \frac{v_p^2}{2g} \qquad (4)$$

where $g = 1.31$ m/s$^2$ is the acceleration due to gravity on Europa. Assuming a particle eruption angle $\theta = 45°$ from the horizontal, the range, $R_p$, which is the distance from the vent that plume particles travel across the surface is:

$$R_p = 2H_p = \frac{v_p^2}{g} \qquad (5)$$

The amount of time that particles spend in these plumes is represented by the particle residence time, $t_R = 2v_p \sin\theta/g$. Assuming a 45° particle eruption angle, $t_R$ is calculated as:

$$t_R = \frac{2v_p \sin\theta}{g} = \frac{v_p\sqrt{2}}{g} \qquad (6)$$

Assuming that dark deposits along lineaments and surrounding lenticulae on Europa are cryoclastic mantlings, *Fagents et al.* [2000] and *Quick et al.* [2013] suggested that plumes required to emplace these deposits would extend, at most, 26 km above the surface. However, recent observations [*Roth et al.*, 2014; *Sparks et al.*, 2016; 2017] suggest that Europa's plumes have maximum heights between 50 and 300 km. In order to account for the broadest suite of plausible plume parameters, we consider eruptions where plume heights range from 1 to 300 km.

*2.1.2 Plume Deposit Parameters*

The volume of a plume deposit can be approximated as that of a thin disk. The area of the plume deposit, as a function of particle size, is:

$$A(r_p) = \pi R_p^2 \tag{7}$$

Consequently, plume deposit thickness as a function of particle size, $T_D(r_p)$, may be expressed by combining equations (3) and (4) from *Quick et al.* [2013]:

$$T_D(r_p) = \frac{4 r_p^3 N_p}{3 R_p^2 (1-\phi)} \tag{8}$$

Here $N_p$ is the total number of particles of a given size in the plume, and hence in the resulting deposit, and $\phi$ represents deposit porosity.

The brightness of Enceladus' interstripe plains, which are believed to be covered by plume fallback, is consistent with that of freshly fallen snow [*Porco et al.*, 2006]. This implies that fresh plume deposits have a snow-like consistency. Freshly fallen snow has 90% pore space so that its porosity, $\phi = 0.9$ [*Cuffey and Paterson*, 2010]. However, any subsequent coalescence and compression of plume particles, perhaps as a result of sintering or other processes, would generate a deposit with a porosity that is more consistent with dense snow, for which $\phi = 0.45$ [*Quick et al.*, 2013]. We assume a slightly higher minimum porosity, $\phi = 0.5$, for plume deposits in which particles have undergone a significant amount of compression and coalescence. We therefore assume that plume deposits have minimum porosities of 0.5, and maximum porosities of 0.9.

The masses of individual plume particles, which must be known in order to apply the model of *Schmidt et al.* [2008] to plumes with $H \geq 50$ km, is defined by $m_p = V_p \rho_p$, where $V_p = \frac{4}{3}\pi r_p^3$ is the volume of a particle. Assuming $\rho_p = 920$ kg/m³, $m_p$ = 4.8 x 10⁻¹⁶ kg, 3.9 x 10⁻¹⁵ kg, 3 x 10⁻¹⁴ kg, and 1 x 10⁻¹³ kg for icy particles with $r_p$ = 0.5, 1, 2, and 3 μm, respectively. The total mass of plume particles, $M$, may be expressed as:

$$M = \sum N_p m_p \tag{9}$$

which, as will be shown in the next section, has been taken to be equal to the total mass of vapor in the plume, $M_v$, multiplied by the plume's ice to vapor ratio so that:

$$M = \sum N_p m_p = M_v * \left(\frac{I}{V}\right) = M_v * \left(\frac{1}{n_f} - 1\right) \tag{10}$$

$N_p$ can be alternatively expressed as:

$$N_p = \frac{M_p}{m_p} \tag{11}$$

where $M_p$ is the total mass of particles of the given size in the plume, which depends on the assumed particle size distribution. For the sake of simplicity, we will here assume that $M_p = M/4$ for all 4 discrete particle sizes, which yields values of $N_p$ that are proportional to $1/r_p^3$. This, along with the non-uniform spacing of the particle sizes, is consistent with the observed particle size distributions observed in Enceladus' plume, which have a differential power-law index between 3 and 4 for micron-sized grains [*Ye et al.*, 2014].

In determining $M_v$ for each plume in (10), we have used the only repeat observation of plumes on Europa, i.e., the 50 km tall plume described in *Sparks et al.* [2016; 2017], as a baseline from which to scale plume mass according to height. The reported column density of the 50 km plume is 1.8 x 10²¹ molecules/m², estimated $M_v$ = 5.4 x 10⁶ kg, and the reported number of water molecules in the plume, $N_w$, is 1.8 x 10³² [*Sparks et al.*, 2017]. This implies an estimated plume area equal to 1 x $10^{11}$ m². Scaling plume column density and area according to plume height for a 1 km tall plume returns a column density of 3.6 x $10^{19}$ molecules/m² and a plume area of 2 x $10^{9}$ m². Multiplying these quantities together returns $N_w$ = 7.2 x $10^{28}$ water molecules total in a 1 km tall plume. From here, $M_v$ may be calculated according to the following relation:

$$M_v = \frac{N_W}{N_A} * m_w \qquad (12)$$

where $N_A$ = 6.022 x $10^{23}$ molecules/mol is Avogadro's number, and $m_w$ = 1.8 x $10^{-2}$ kg/mol is the molar mass of one water molecule. Application of (12) returns $M_v$ = 2.2 x $10^{3}$ kg of water vapor for a 1 km tall plume. $M_v$ for plumes with $H$ = 10-300 km are listed in Table 3.

**2.2 Spectral Model**

The computed ranges and deposit thicknesses for various particle sizes can be translated into predictions for deposit spectra using light-scattering models that provide analytical expressions for the wavelength-dependent brightness of a surface [*Hapke,* 1981;1993; *Cuzzi and Estrada,* 1998; *Shkuratov et al.,* 1999]. These model spectra depend on both the composition and texture of the regolith, which are often quantified in terms of the product $\alpha S$, where $S$ is the mean scattering length of the photons within the surface regolith (also known as the regolith's "grain size"), and $\alpha$ is the absorption coefficient of the plume material, which is given by the expression $\alpha = 4\pi\kappa/\lambda$. Here $\kappa_r$ is the imaginary part of the material's refractive index and $\lambda$ is the wavelength of the radiation.

For this particular analysis, we will assume that the deposits are sufficiently thick so that the underlying material does not contribute to the spectrum. For the near-infrared wavelengths considered here, this corresponds to a deposit at least a few tens of microns thick, which is reasonable for the sources considered above. Finally, we assume that the scattering length $S$ is equal to the average grain radius, $r_{eff}$, in the deposit, which is a function of the distance from the vent. For the sake of simplicity, we assume that the deposit is composed of pure water ice. While plume deposits could include non-ice materials with distinctive spectral signatures, at present there are few constraints on the nature or concentration of such contaminants. Hence for this initial study we have chosen to focus on spectral trends associated with variations in the average grain size of icy particles in the plume deposits. Assuming a fixed composition allows us to utilize the optical constants determined by *Mastrapa et al.* [2009], specifically values of $\kappa_r$ at each wavelength for crystalline ice at 120 K. The above assumptions will enable us to obtain a qualitative sense of the spectral trends in plume deposits. However, future work that considers a range of particle compositions and full particle size distributions will be needed to derive robust estimates of certain quantities such as the depths of specific bands.

In practice, we compute $r_{eff}$ by first interpolating the above ranges and deposit thicknesses onto a regular grid of 100 particle sizes between 0.5 and 3 $\mu$m. Since these parameters are roughly power-law functions of the particle size, we perform these interpolations on the logarithms of the relevant parameters (i.e., we take the logarithm of the particle sizes and ranges, interpolate linearly between the computed values, and then take the exponential to recover the interpolated ranges). Then, for each radial distance from the vent, we compute the average particle size, $r_{eff}$, as the weighted average of the particle sizes in the deposit using the following formula:

$$r_{eff} = \frac{\sum r_p T_D(r_p)}{\sum T_D(r_p)} \tag{13}$$

where $r_p$ are the individual particle sizes and $T_D$ are the deposit thicknesses. Note that this sum only considers particles between 0.5 and 3$\mu$m, so the particle size distribution emerging from the vent is assumed to have hard cutoffs at 0.5 and 3$\mu$m.

The above estimates of $\alpha$ and $r_{eff}=S$ can be used to compute the predicted spectra at each radius using the analytical models from *Cuzzi and Estrada* [1998], which uses a Hapke-based formalism, or *Shkuratov et al.* [1999]. In practice, these two papers provide very different formulas for albedo as a function of $\alpha S$, and it is well known that the Hapke and Shkuratov scattering theories can yield different estimates of the composition and effective scattering lengths required to match a given spectrum [*Poulet et al.*, 2002]. In part, this is because *Cuzzi and Estrada* [1998] compute the albedo of regolith, while *Shkuratov et al.*[1999] compute the albedo for a one-dimensional model of a regolith surface. For this analysis, we prefer to use the *Shkuratov et al.* [1999] model because it is explicitly designed for spectral analysis, while Hapke-based models are better optimized for photometric studies. Hence for this analysis the expected value of the deposit's brightness for a given $\alpha S$ is computed using equations 8-12 from *Shkuratov et al.* [1999]. For the sake of simplicity, we assume here that the real part of the grains' refractive index is $n = 1.3$ (appropriate for ice-rich material) and zero porosity (note that including a finite porosity changes the overall strength of spectral features, but not the trends with distance from the vent). In this case, the relevant formula for the brightness can be written as:

$$A_1 = \frac{1+r_b^2-r_f^2}{2r_b} - \sqrt{\left(\frac{1+r_b^2-r_f^2}{2r_b}\right)^2 - 1} \qquad (14)$$

where the parameters $r_b$ and $r_f$ are given by the expressions:

$$r_b = R_b + \frac{1}{2}\frac{(1-R_e)(1-R_i)e^{-2\alpha S}}{1-R_i e^{-\alpha S}}$$

$$r_f = (R_e - R_b) + (1-R_e)(1-R_i)e^{-\alpha S} + \frac{1}{2}\frac{(1-R_e)(1-R_i)e^{-2\alpha S}}{1-R_i e^{-\alpha S}},$$

and the coefficients $R_i$, $R_e$ and $R_b$ are set by our choice of $n$:

$R_i \approx 1.04 - 1/n^2 \approx 0.45$

$R_e \approx (n-1)^2/(n+1)^2 + 0.05 \approx 0.067$

$R_b \approx (0.28\, n - 0.20)\, R_e \approx 0.011$

When using these formulae, it is important to understand that the parameter derived by *Shkuratov et al.* [1999] (here denoted $A_1$) is a ``brightness coefficient'' of a one-dimensional model system viewed at low phase angles [*Shkuratov et al.* 1999]. The value of $A_1$ at any given wavelength should therefore not be mistaken for the Bond or single-scattering albedo of the surface [*Hedman et al.* 2013]. It is also important to note that the simplifications associated with the above model will fail around strong water-ice absorption bands, where the real index deviates strongly from 1.3 and the imaginary index is large. This model therefore does not provide reliable information about the shape of the deep water-ice absorption band around 3 microns. However, as prior work demonstrates that this model can reproduce the overall shape and depths of the 1.5 and 2.0 $\mu$m bands quite well for ice-rich surfaces [*Hedman et al.* 2013], this model is adequate for this initial study.

Once we have computed the model spectra of the plume deposits, we may calculate spectral parameters such as the 1.5 and 2.0 $\mu$m band depths. These band depths are simply the difference in brightness between the center of the band and the continuum on either side, normalized to the continuum brightness level. For the 1.5 $\mu$m band depth, we use the average $A_1$ between 1.5 and 1.55 $\mu$m to define the brightness in the center of the band, while the average $A_i$ between 1.35-1.40 $\mu$m and 1.8-1.85 $\mu$m defines the continuum brightness level. For the 2.0 $\mu$m band depth, the average $A_1$ between 2.00 and 2.05 $\mu$m defines the brightness in the center of the band, while the average $A_i$ between 1.80-1.85 $\mu$m and 2.20-2.25 $\mu$m defines the continuum brightness level. These simple estimates of the band depths are sufficient to illustrate trends in the deposit's spectral parameters with distance from the vent.

**Table 2.** Plume velocity ($v$) and height ($H$) as a function of wt% of water vapor ($n_{gas}$). Gas mass fraction, $n_f = n_{gas}/100$

| $n_{gas}$ (%) | $v$ (m/s) | $H$ (km) |
|---|---|---|
| 0.1 | 30 | 0.34 |
| 0.2 | 42 | 0.67 |
| 0.3 | 52 | 1.0 |
| 0.4 | 60 | 1.3 |
| 0.5 | 67 | 1.7 |
| 0.6 | 73 | 2.0 |
| 0.7 | 79 | 2.3 |
| 0.8 | 84 | 2.7 |
| 0.9 | 89 | 3.0 |
| 1 | 94 | 3.4 |
| 2 | 133 | 6.7 |
| 3 | 163 | 10.1 |
| 4 | 188 | 13.4 |
| 5 | 210 | 16.8 |
| 6 | 230 | 20 |
| 7 | 249 | 23 |
| 8 | 266 | 27 |
| 9 | 282 | 30 |
| 10 | 298 | 34 |
| 15 | 364 | 50.3 |
| 20 | 421 | 67 |
| 25 | 471 | 84 |
| 30 | 515 | 101 |
| 35 | 557 | 117 |
| 40 | 595 | 134 |
| 45 | 631 | 151 |
| 50 | 665 | 168 |
| 55 | 698 | 185 |
| 60 | 729 | 201 |
| 65 | 759 | 218 |
| 70 | 787 | 235 |
| 75 | 815 | 252 |
| 80 | 842 | 268 |
| 85 | 868 | 285 |
| 90 | 893 | 302 |
| 95 | 917 | 319 |
| 100 | 941 | 335 |

**Table 3.** Cryovolcanic Plume Parameters: Plume Height ($H$), wt% ($n_{gas}$) and mass fraction ($n_f$) of water vapor, total mass of water vapor in the plume ($M_v$), total mass of icy particles in the plume ($M$), and ice to vapor ratios ($I/V$)

| H (km) | $n_{gas}$ (%) | $n_f$ | $M_v$ (kg) | M (kg) | I/V |
|---|---|---|---|---|---|
| 1 | 0.3 | 0.003 | 2.2 x 10³ | 7.2 x 10⁵ | 332 |
| 10 | 3.0 | 0.030 | 2.2x 10⁵ | 6.9 x 10⁶ | 32 |
| 25 | 7.4 | 0.074 | 1.4 x 10⁶ | 1.7 x 10⁷ | 12.5 |
| 50 | 15 | 0.15 | *5.4 x 10⁶ | 3.1 x 10⁷ | 5.7 |
| 100 | 30 | 0.30 | 2.2 x 10⁷ | 5 x 10⁷ | 2.3 |
| 200 | 60 | 0.60 | 8.6 x 10⁷ | 5.8 x 10⁷ | 0.67 |
| 300 | 90 | 0.90 | 2 x 10⁸ | 2 x 10⁷ | 0.1 |

*Extracted from *Sparks et al.* [2017]

## 3. Results

### 3.1 Dynamical Model Results

*3.1.1. Plumes with H = 1 km*

According to (1) and (4), eruptions with a 0.3 wt% water vapor content will produce plumes that extend 1 km above Europa's surface. In this case, $v_{gas}$ = 52 m/s, and the mass fraction of gas, $n_f = n_{gas}/100 = 0.003$ (Table 2). A 0.3 wt% water vapor content means that 99.7wt % or .997 mass fraction of a 1 km tall plume consists of icy particulates. In this case, $I/V$ = .997/.003 = 332 (Table 3). Substituting $M_v$ = 2.2 x 10³ kg and $I/V$ = 332 into (10) returns $M$ = 7.2 x 10⁵ kg for the total mass of ice in the plume. If we assume that the total mass of particles of each size is equal to ¼ the total mass of ice in the plume, then $M_p$ = ¼ $M$ = 1.8 x 10⁵ kg as the total mass of particles of each size (i.e., $r_p$ = 0.5 μm, 1 μm, etc.) in the plume, as well as in the resulting deposit. As mentioned in Section 2.1, icy particles in a 1 km tall plume will remain dispersed in a dense column of gas near the vent and will not be reaccelerated during the course of the eruption [*Yeoh et al.*, 2015]. Hence, we can assume that particles in these small plumes will travel at maximum speeds close to the gas speed (cf. *Fagents et al.* [2000]). In addition, the maximum particle deposition radius will be ~ 2 km from the vent (Table 4a).

Based on the duration of observations during which *Sparks et al.* [2017] identified plumes on Europa, those authors suggested that Europa's plumes have ~ 1 hour eruption timescales. Thus, assuming continuous eruptions occur for at least an hour, deposits with 50% porosity ($\phi = 0.5$) would accumulate at rates of 8.6 x $10^{-9}$ m/s, while those with 90% porosity ($\phi = 0.9$) would accumulate at rates of 4.3 x $10^{-8}$ m/s (Table 4a). At these deposition rates, it would take almost 4 years to produce a 1 m thick deposit with 50% porosity; a similar deposit with 90% pore space would take just under 9 months to form. In both cases the deposit would be spread over an area of 12.6 km² on the surface (Table 4a). The time for 10 m thick deposits to form can be determined by multiplying the time it takes for 1 m deposits to accumulate by a factor of 10, so that a 10 m thick deposit with 50% porosity would take 37 years to form, while a similar deposit with 90% porosity would accumulate in just over 7 years (Table 4a). According to (6), particles in a 1 km tall plume would have a residence time, $t_R$ of 55 sec.

In order to determine the maximum distance that icy particles will travel across the surface, we have considered 45° as the eruption angle at which plume particles will be ejected. In this way we are able to obtain the maximum deposit radius for particles of a certain size. We assume that all particles will be uniformly emplaced within a circle for which the outer radius is commensurate with their maximum travel distance across the surface. However, it is likely that particles will be ejected from the plume at a range of initial angles between 1° and 90° from the horizontal [*Fagents et al.*, 2000; *Glaze and Baloga*, 2000; *Quick et al.*, 2013], so that there will be overlap between deposits whose constituent particles are primarily of one size. In other words, plume deposits consisting primarily of particles with $r_p = 0.5$ μm may overlap with deposits whose constituent particles are mostly 2 or 3 μm in radius. It is therefore likely that deposits containing variable particle sizes will build up on the surface. Assuming that each deposit contains particles of multiple sizes, we find that a surface deposit emplaced during a single eruption of a 1 km tall plume could be ~ 0.12 mm thick if the resulting deposit is 50% porous, and 0.62 mm thick if the deposit has 90% porosity (Table 4a).

### 3.1.2. Plumes with H = 25 km

These plumes would have 256 m/s gas speeds and would contain 7.4 wt% water vapor and 92.6 wt% icy particles, resulting in an $I/V$ = 12.5 (Tables 2 & 3). Utilizing (9)-(12), and scaling plume mass with height as described in Section 2.1, returns $M_v$ = 1.4 x$10^6$, $M$ = 1.7 x $10^7$ kg and $M_p$ = 4 x $10^6$ kg. Here, the maximum particle deposition radius is 50 km (Table 4a). The resulting plume deposits would be spread over a relatively wide area of Europa's surface, and particle deposition rates would be 3 x $10^{-10}$ m/s for deposits with 50% porosity and 1.6 x $10^{-9}$ m/s for deposits with 90% pore space (Table 4a). Total deposits thicknesses, per eruption, would be 5 x $10^{-3}$ mm for deposits with 50% pore space, and ~ 0.02 mm for deposits with 90% pore space. In the case of deposits that are 90% porous, 1 m thick deposits will take less than 20 years to accumulate. Conversely, 1 m thick deposits with 50% porosity would take almost 100 years to form (Table 4a).

### 3.1.3. Plumes with H = 50 km

Plumes that extend 50 km above Europa's surface would be composed of ~ 15 wt% water vapor and 85 wt % icy particles. In this case $v_{gas}$ = 362 m/s and $I/V$ = 5.7 (Tables 2 - 3). *Sparks et al.* [2017] report observations of a 50 km tall plume on Europa, with an estimated water vapor content of 5.4 x $10^6$ kg. Substituting $M_v$ = 5.38 x$10^6$ kg and $I/V$ = 5.7 into (10) returns $M$ = 3.1 x $10^7$ kg and $M_p$ = 7.7 x $10^6$ kg. As previously mentioned, the dynamics of particles in plumes with $H \geq 50$ km may be described by equations (2) and (3). Employing (3) and assuming that plume expansion begins at $T$ = 240 K, $\rho_{gas}$ = 4.85 x $10^{-3}$ kg/m³, $\beta$ = 0.2, and $L$ = 0.1 m, commensurate with the minimum collision length of plume particles on Enceladus, [*Schmidt et al.*, 2008], returns $r_c$ = 1 μm. However, all particles peak at speeds less than $v_{gas}$ (Table 4b), and only cryoclastic particles with $r_p$ = 0.5$\mu$m will reach the maximum particle deposition radius of 64 km (Fig. 1). The areal extent of deposits emplaced by a 50 km tall plume would be quite broad and could cover an area up to 1.3 x $10^4$ km$^2$ (Fig. 2 & Table 4b). Particle deposition rates for hour-long eruptions range from 4 x $10^{-10}$ – 6 x$10^{-9}$ m/s, with the highest deposition rates occurring for particles with $r_p$ = 3$\mu$m (Table 4b). 1m thick deposits with 50% pore space could take as little as 6 years to accumulate if primarily composed of particles with $r_p$ = 3$\mu$m, while 1m thick deposits composed of particles with $r_p$ = 2$\mu$m could form in approximately 14 years. Conversely, 1 m thick deposits with 50% pore space could take in excess of 40 years to accumulate if primarily composed of particles with $r_p$ = 1 $\mu$m, and 89 years to accumulate if primarily composed of particles with $r_p$ = 0.5 $\mu$m (Table 4b). For all particle sizes considered, 1 m thick deposits with 90% pore space would take, at most, 20 years to form. Of note is that 1 m thick deposits, with $\phi$ = 0.9, that are primarily composed of particles with $r_p$ = 1-2 $\mu$m would take 9 and 3 years to accumulate, respectively, while it would only take deposits composed of particles with $r_p$ = 3 $\mu$m a year to form a 1 m thick layer on the surface. If deposits consist of particles that vary in size so that particles with $r_p$ ranging from 0.5 -3$\mu$m are present, total deposit thicknesses per eruption would be 0.03 mm and 0.16 mm for deposits with $\phi$ = 0.5 and 0.9, respectively.

### 3.1.4. Plumes with H = 200 km

Plumes that extend 200 km above Europa's surface would be composed of ~ 60 wt% water vapor and 40 wt% icy particles (Table 2). In this case, $v_{gas}$ = 724 m/s, as previously reported [*Roth et al.*, 2014] and $I/V$ = 0.67 (Table 3). We find that $M_v$ would be 8.6 x $10^7$ kg for plumes with $H$ = 200 km. Substituting this value for $M_v$ and $I/V$ = 0.67 into (10) returns $M$ = 5.8 x $10^7$ kg, and $M_p$ = 1.4 x $10^7$ kg. Although according to (3) $r_c$ = 0.5 $\mu$m, all particle velocities peak at values that are substantially less than the gas speed. Particle deposition rates would be between $10^{-11}$ and $10^{-9}$ m/s for deposits with 50% porosity, and between $10^{-10}$ and $10^{-8}$ m/s for deposits with 90% porosity. (Table 4b). Total deposit thicknesses, per eruption, are 0.02 mm and 0.11 mm for deposits with $\phi$ = 0.5 and 0.9, respectively. Our analysis shows that for the case of such large plumes, 1m thick deposits with 50% pore space would take between 7.5 and 24 years to accumulate if primarily composed of larger particles with $r_p$ = 2-3 $\mu$m, and on the order of 100 years to accumulate if composed of smaller particles with $r_p$ = 0.5-1 $\mu$m (Table 4b). 1m thick deposits with 90% pore space would take $\leq$ 24 years to accumulate if primarily composed of particles with $r_p \geq$ 1 $\mu$m. However if primarily composed of particles with $r_p$ = 0.5 $\mu$m, 1m thick deposits could take as much as 74 years to form. Particles ejected by plumes this size could be deposited as far as 180 km from their eruptive source regions (Fig. 1), and depending on particle size, the resulting deposits could be spread over areas as large as 101,510 km$^2$ (Fig. 2 & Table 4b).

For plumes with $H \geq$ 50 km, we find that only particles with submicron radii travel at speeds identical to the gas speed (Fig. 3). This is similar to the case for plumes on Enceladus [*Schmidt et al.*, 2008]. Utilizing equations (1)-(3), we find that for plumes with $H$ = 50, 100, and 200 km, only particles with $r_p \leq$ 2 x 10$^{-3}$ $\mu$m, 1 x 10$^{-3}$ $\mu$m, and 7 x 10$^{-4}$ $\mu$m respectively, would travel at speeds identical to the gas speed (Fig. 3). In the case of 300 km tall plumes, only particles with $r_p \leq$ 6 x 10$^{-5}$ $\mu$m would travel at the gas speed (Fig. 3). Hence the larger the plume, the smaller the particles must be in order to attain the gas speed. Figure 4 illustrates that in all of the plumes considered in our study, the population of small particles ($r_p$ = 0.5 $\mu$m) is larger than the population of large particles ($r_p$ = 3 $\mu$m) by 2-3 orders of magnitude. Additional details regarding particle deposition for plumes for plumes with $H$ = 10, 100 and 300 km can be found in Tables 4a-c.

**Table 4a.** Plume and Deposit Parameters for Small Plumes

|  | $H = 1$ km<br>$v = 51$ m/s | $H = 10$ km<br>$v = 162$ m/s | $H = 25$ km<br>$v = 256$ m/s |
|---|---|---|---|
| $v_p$ (m/s) | 51 | 162 | 256 |
| $H_p$ (km) | 1 | 10 | 25 |
| $R_p$ (km) | 2 | 20 | 50 |
| $t_R$ (s) | 55 | 175 | 276 |
| $A$ (km$^2$) | 12.6 | 1257 | 7854 |
| $^a D_{\phi=50\%}$ (m/s) | 8.6 x 10$^{-9}$ | 8 x 10$^{-10}$ | 3 x 10$^{-10}$ |
| $T_{D_{\phi=50\%}}$ (mm) | 0.03 | 0.003 | 1.2 x 10$^{-3}$ |
| $T_{total\,\phi=50\%}$ per eruption | **0.12 mm** | **0.01 mm** | **0.005 mm** |
| $^b t_{deposit} = 1$ m | 3.7 yr | 38 yr | 98 yr |
| $^c t_{deposit} = 10$ m | 37 yr | 383 yr | 976 yr |
| $^d D_{\phi=90\%}$ (m/s) | 4.3 x 10$^{-8}$ | 4.1 x 10$^{-9}$ | 1.6 x 10$^{-9}$ |
| $T_{D_{\phi=90\%}}$ (mm) | 0.15 | 0.015 | 5.8 x 10$^{-3}$ |
| $T_{total\,\phi=90\%}$ per eruption | **0.62 mm** | **0.06 mm** | **0.02 mm** |
| $^e t_{deposit} = 1$ m | 0.74 yr | 7.7 yr | 19.5 yr |
| $^f t_{deposit} = 10$ m | 7.4 yr | 77 yr | 195 yr |

$^{a,d}$Particle deposition rate assuming each individual eruption is continuous for 1 hour
$^{b,c}$time to accumulate 1 m and 10 m thick deposits assuming individual eruptions last for one hour, when $\phi = 50\%$
$^{e,f}$time to accumulate 1 m and 10 m thick deposits assuming individual eruptions last for one hour, when $\phi = 90\%$

**Table 4b.** Plume and Deposit Parameters for Large Plumes

|  | $H = 50$ km<br>$v = 362$ m/s | | | | $H = 100$ km<br>$v = 512$ m/s | | | | $H = 200$ km<br>$v = 724$ m/s | | | |
|---|---|---|---|---|---|---|---|---|---|---|---|---|
|  | $r_p$ ($\mu$m) | | | | $r_p$ ($\mu$m) | | | | $r_p$ ($\mu$m) | | | |
|  | 0.5 | 1 | 2 | 3 | 0.5 | 1 | 2 | 3 | 0.5 | 1 | 2 | 3 |
| $v_p$ (m/s) | 291 | 243 | 182 | 146 | 380 | 302 | 214 | 166 | 485 | 365 | 244 | 183 |
| $H_p$ (km) | 32 | 22 | 13 | 8 | 55 | 35 | 17 | 10 | 90 | 51 | 23 | 13 |
| $R_P$ (km) | 64 | 45 | 25 | 16 | 110 | 70 | 35 | 21 | 180 | 102 | 45 | 26 |
| $t_R$ (s) | 314 | 262 | 197 | 158 | 410 | 326 | 231 | 179 | 524 | 394 | 263 | 198 |
| $A$ (km$^2$) | 13,043 | 6344 | 2029 | 837 | 38,078 | 15,200 | 3843 | 1384 | 101,510 | 32,466 | 6483 | 2062 |
| $^a D_{\phi=50\%}$ (m/s) | 4x10$^{-10}$ | 7 x 10$^{-10}$ | 2 x 10$^{-9}$ | 6 x 10$^{-9}$ | 2 x10$^{-10}$ | 5 x10$^{-10}$ | 2 x10$^{-9}$ | 5 x10$^{-9}$ | 9 x 10$^{-11}$ | 3 x 10$^{-10}$ | 1 x10$^{-9}$ | 4 x10$^{-9}$ |
| $T_{D_{\phi=50\%}}$ (mm) | 1 x 10$^{-3}$ | 3 x 10$^{-3}$ | 0.01 | 0.02 | 7 x 10$^{-4}$ | 2 x 10$^{-3}$ | 7 x 10$^{-3}$ | 0.02 | 3 x 10$^{-4}$ | 1 x 10$^{-3}$ | 5 x 10$^{-3}$ | 0.02 |
| $T_{total\,\phi=50\%}$ per eruption | **0.03 mm** | | | | **0.03 mm** | | | | **0.02 mm** | | | |
| $^b t_{deposit} = 1$ m | 89 yr | 43 yr | 14 yr | 6 yr | 161 yr | 64 yr | 16 yr | 6 yr | 369 yr | 118 yr | 24 yr | 7.5 yr |
| $^c t_{deposit} = 10$ m | 891 yr | 433 yr | 139 yr | 57 yr | 1614 yr | 644 yr | 163 yr | 59 yr | 3689 yr | 1180 yr | 236 yr | 75 yr |
| $^d D_{\phi=90\%}$ (m/s) | 2 x10$^{-9}$ | 4 x10$^{-9}$ | 1 x 10$^{-8}$ | 3 x 10$^{-8}$ | 9.8 x10$^{-10}$ | 2 x10$^{-9}$ | 9.7 x10$^{-9}$ | 3 x 10$^{-8}$ | 4 x 10$^{-10}$ | 1 x 10$^{-9}$ | 7 x10$^{-9}$ | 2 x10$^{-8}$ |
| $T_{D_{\phi=90\%}}$ (mm) | 6 x 10$^{-3}$ | 0.01 | 0.04 | 0.1 | 4 x 10$^{-3}$ | 9 x 10$^{-3}$ | 0.03 | 0.1 | 2 x 10$^{-3}$ | 5 x 10$^{-3}$ | 0.02 | 0.08 |
| $T_{total\,\phi=90\%}$ per eruption | **0.16 mm** | | | | **0.14 mm** | | | | **0.11 mm** | | | |
| $^e t_{deposit} = 1$ m | 18 yr | 9 yr | 3 yr | 1 yr | 32 yr | 13 yr | 3 yr | 1 yr | 74 yr | 24 yr | 5 yr | 1.5 yr |
| $^f t_{deposit} = 10$ m | 178 yr | 87 yr | 28 yr | 11 yr | 323 yr | 129 yr | 33 yr | 12 yr | 738 yr | 236 yr | 47 yr | 15 yr |

$^{a,d}$Particle deposition rate assuming each individual eruption is continuous for 1 hour
$^{b,c}$time to accumulate 1 m and 10 m thick deposits assuming individual eruptions last for one hour, when $\phi = 50\%$
$^{e,f}$time to accumulate 1 m and 10 m thick deposits assuming individual eruptions last for one hour, when $\phi = 90\%$

**Table 4c**. Plume and Deposit Parameters for Plumes with $H = 300$ km

|  | $H = 300$ km $v = 887$ m/s | | | |
|---|---|---|---|---|
|  | $r_p$ ($\mu$m) | | | |
|  | 0.5 | 1 | 2 | 3 |
| $v_p$ (m/s) | 553 | 402 | 260 | 192 |
| $H_p$ (km) | 117 | 62 | 26 | 14 |
| $R_P$ (km) | 234 | 123 | 52 | 28 |
| $t_R$ (s) | 597 | 434 | 281 | 207 |
| $A$ (km$^2$) | 171,520 | 47,852 | 8366 | 2494 |
| [a]$D_{\phi=50\%}$ (m/s) | $1.7 \times 10^{-11}$ | $6 \times 10^{-11}$ | $4 \times 10^{-10}$ | $1 \times 10^{-9}$ |
| $T_{D_{\phi=50\%}}$ (mm) | $6 \times 10^{-5}$ | $2 \times 10^{-4}$ | $1 \times 10^{-3}$ | $4 \times 10^{-3}$ |
| $T_{total_{\phi=50\%}}$ per eruption | $5.86 \times 10^{-3}$ mm | | | |
| [b]$t_{deposit} = 1$ m | 1825 yr | 509 yr | 89 yr | 27 yr |
| [c]$t_{deposit} = 10$ m | 18,250 yr | 5092 yr | 890 yr | 265 yr |
| [d]$D_{\phi=90\%}$ (m/s) | $9 \times 10^{-11}$ | $3 \times 10^{-10}$ | $2 \times 10^{-9}$ | $6 \times 10^{-9}$ |
| $T_{D_{\phi=90\%}}$ (mm) | $3 \times 10^{-4}$ | $1 \times 10^{-3}$ | $6 \times 10^{-3}$ | $2 \times 10^{-2}$ |
| $T_{total_{\phi=90\%}}$ per eruption | 0.03 mm | | | |
| [e]$t_{deposit} = 1$ m | 365 yr | 102 yr | 18 yr | 5 yr |
| [f]$t_{deposit} = 10$ m | 3650 yr | 1018 yr | 178 yr | 53 yr |

[a,d]Particle deposition rate assuming each individual eruption is continuous for 1 hour
[b,c]time to accumulate 1 m and 10 m thick deposits assuming individual eruptions last for one hour, when $\phi = 50\%$
[e,f]time to accumulate 1 m and 10 m thick deposits assuming individual eruptions last for one hour, when $\phi = 90\%$

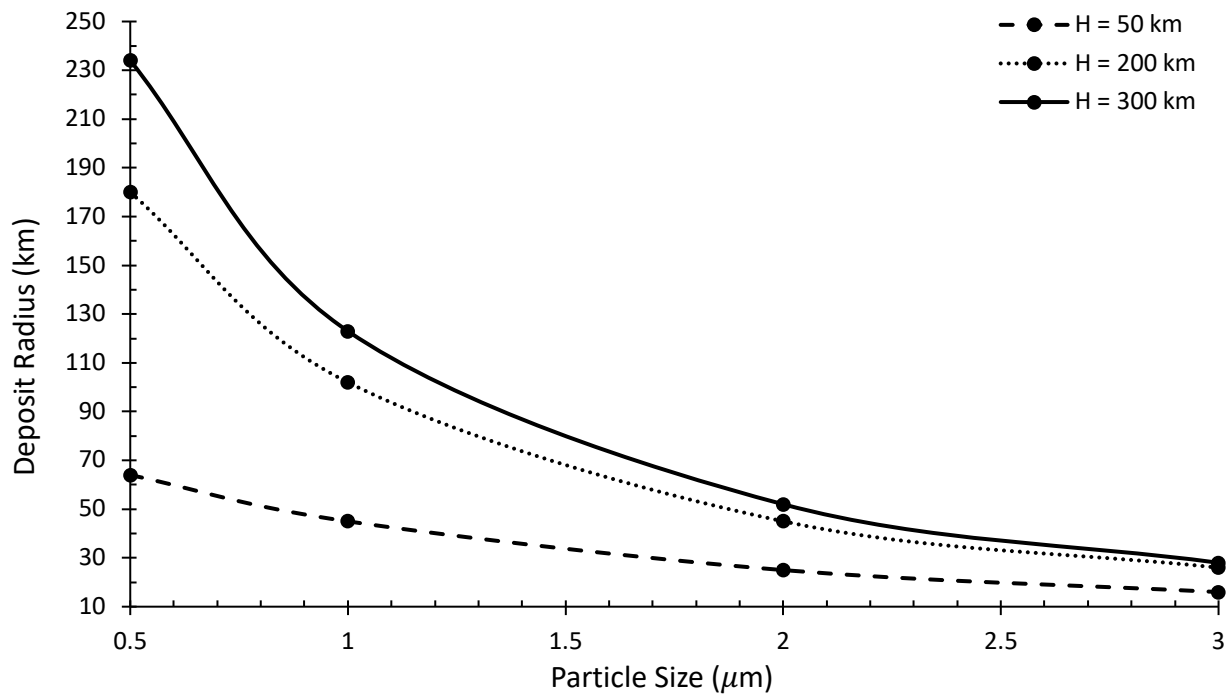

**Figure 1.** Maximum distance beyond the vent that icy particles reach, as a function of particle size, for plumes that are 50 km [*Sparks et al.*, 2017] and 200 km [*Roth et al.*, 2014] tall. We also plot maximum particle range for 300 km tall plumes, which may represent the largest possible plume observed on Europa according to *Roth et al.* [2014]. Each black dot, from the top left to the lower right of each data series, represents the particle sizes considered in this analysis: $r_p$ = 0.5 $\mu$m, 1 $\mu$m, 2 $\mu$m, and 3 $\mu$m.

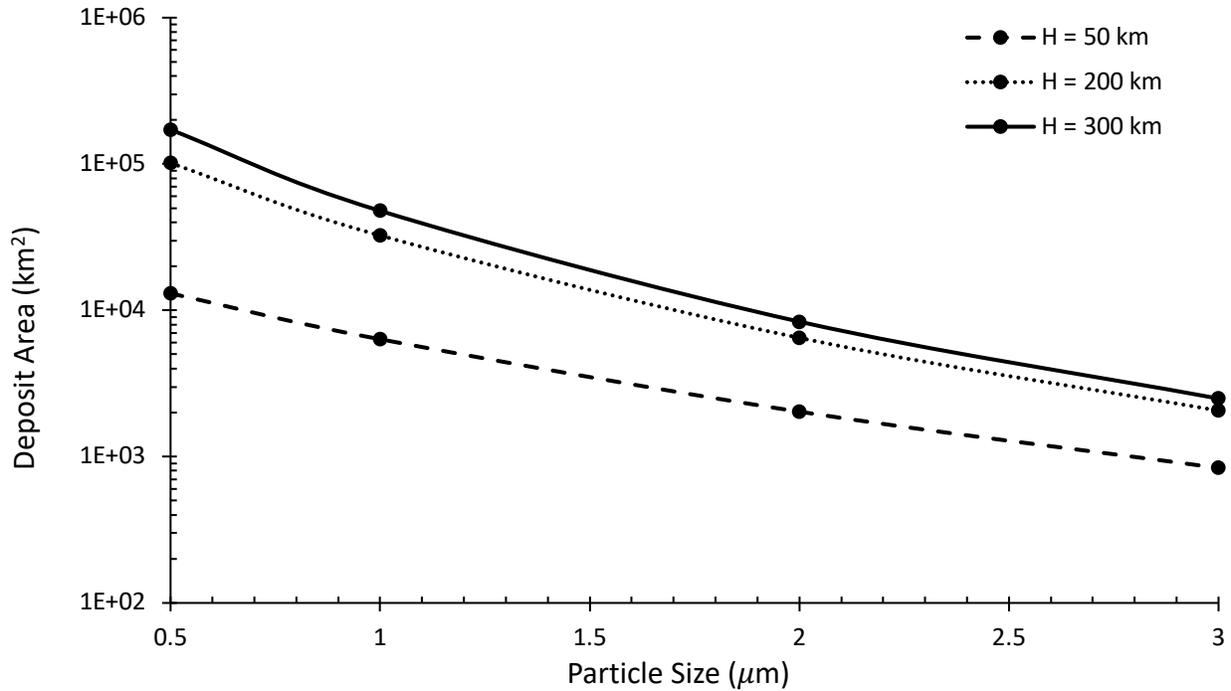

**Figure 2.** Areal extent of plume deposits as a function of primary particle size for plumes that are 50 km [*Sparks et al.*, 2017], 200 km, and 300 km tall [*Roth et al.*, 2014]. Each black dot, from the top left, to the lower right of each data series, represents the particle sizes considered in this analysis: $r_p$ = 0.5 $\mu$m, 1 $\mu$m, 2 $\mu$m, and 3 $\mu$m.

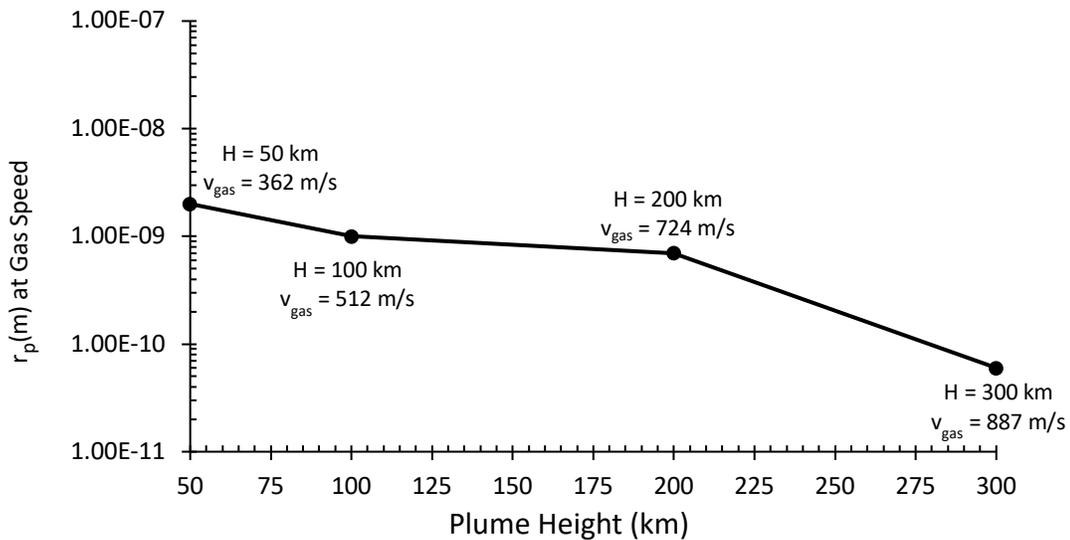

**Figure 3.** Size of particles that attain the gas speed as a function of plume height. The larger the plume, and the higher the gas velocity, the smaller constituent particles must be to remain entrained with the gas. For plume heights greater than 100 km, particle radii must be ~ $\leq 10^{-10}$ m to attain the gas speed.

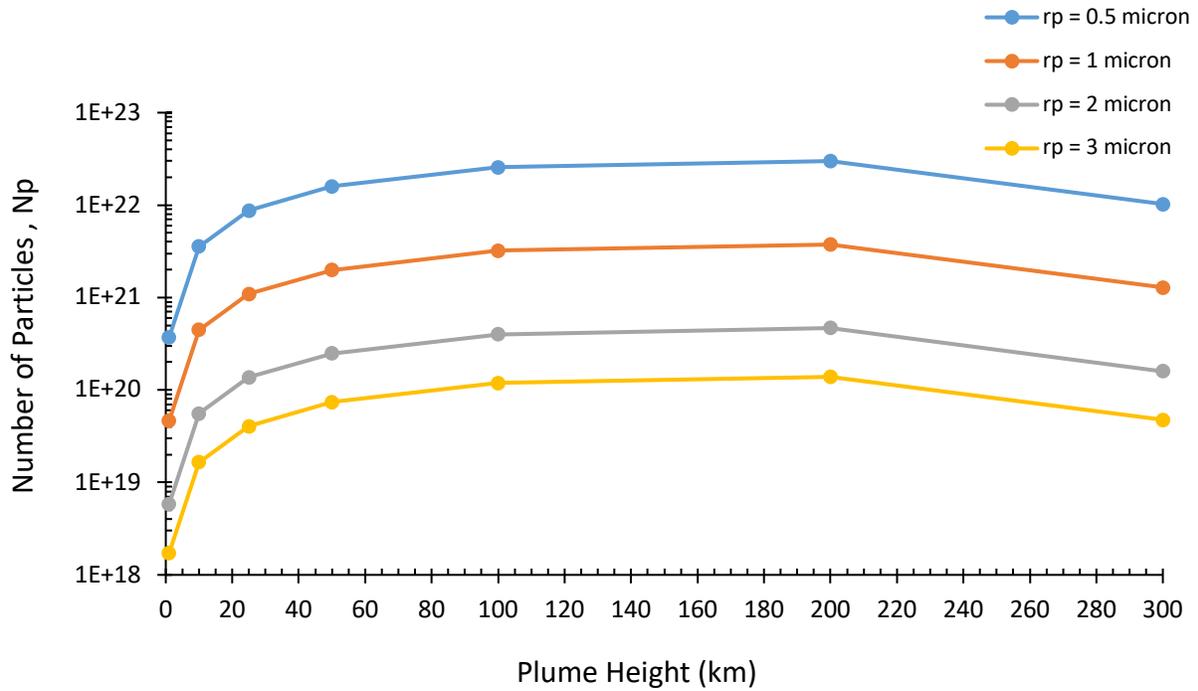

**Figure 4.** Particle population as a function of plume height. Regardless of plume height, the number of particles in the smallest particle population in the plume ($r_p = 0.5\ \mu$m) is larger than the number of particles in the biggest particle population in the plume ($r_p = 3\ \mu$m) by 2-3 orders of magnitude.

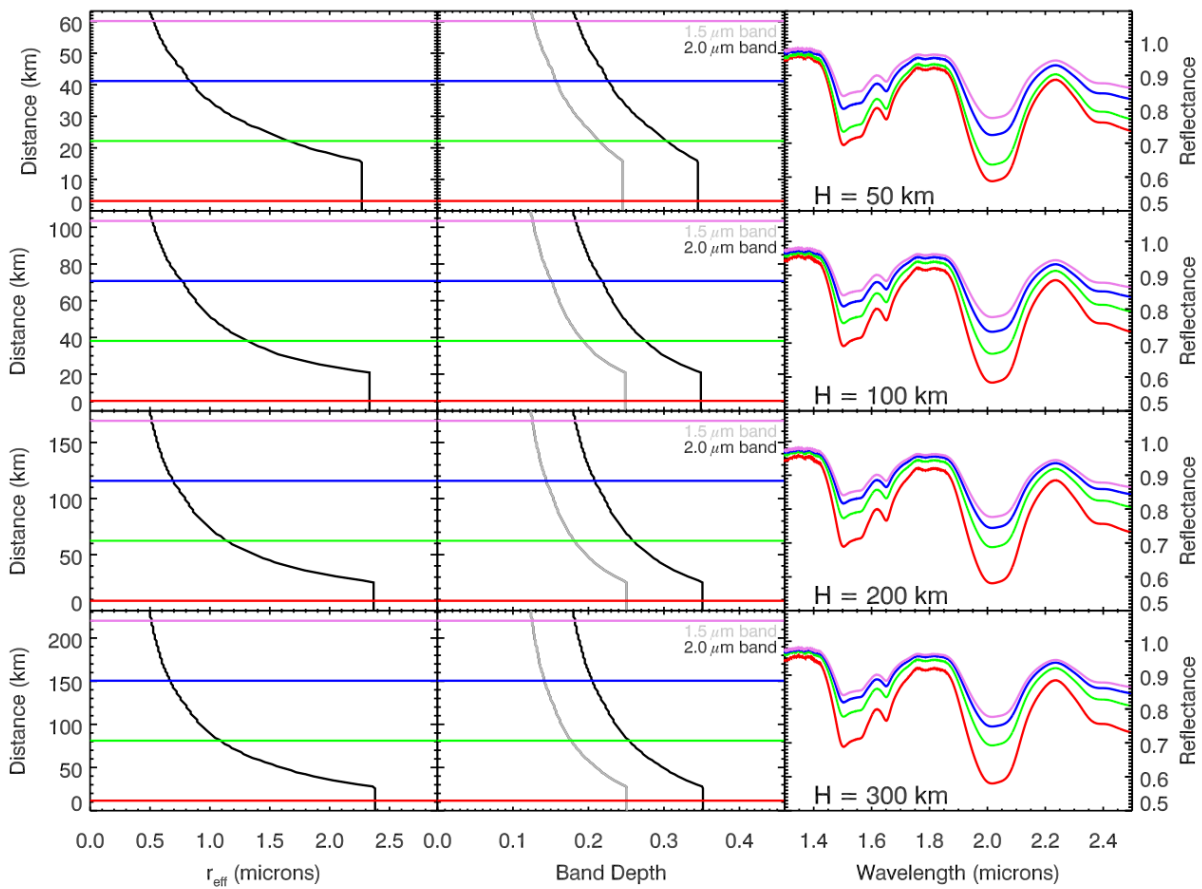

**Figure 5.** Average effective particle size and model spectra for deposits associated with a 50-300 km tall plumes. The left panels show average particle size as a function of distance from the vent, while the middle panel shows the expected depths of the water-ice bands as a function of distance from the vent. The right panel shows representative spectra at the distances marked with horizontal lines in the previous panels. Note that while the scale of the plumes changes, the overall trends in spectral properties are remarkably similar.

## 3.2 Spectral Model Results

Figure 5 illustrates how the average effective particle size and model surface spectra vary with distance from the vent for plumes with heights greater than 50 km (for shorter plumes, the uniform particle velocity leads to deposits with uniform spectral properties). The left panels show the effective average particle size versus distance from the vent. The middle panels show how the depths of the 1.5 and 2.0 $\mu$m water ice bands vary with distance from the vent, and the right panels show representative spectra between 1.3 and 2.5 $\mu$m at a few selected distances from the vent. Recall that all these plots assume that the deposit is at least a few tens of microns thick, so that the underlying terrain does not contribute to the spectra.

For all of the simulated plumes, the effective particle size decreases with distance from the vent, which gives rise to a corresponding reduction in the depths of the water ice bands. This basic result simply reflects the fact that smaller particles are launched at higher speeds and so reach larger distances from the vent. Indeed, the observed trends are very similar for all the plume deposits shown in Figure 5. However, it is worth noting that the detailed shape of these trends is different for each plume. In particular, for the smaller plumes, the effective average particle size and band depths are constant over a larger fraction of the deposit because the particles are all launched at nearly the same speed (cf. Table 4a). For taller plumes, there are more substantial variations in the deposit's effective average particle size and band depths with distance from the vent. In addition, the maximum effective average particle sizes and band depths near the vent are slightly larger than they are for more compact plumes. This happens because the dispersion of launch velocities is higher, so the larger particles are now more concentrated near the vent.

## 4. Discussion

While the above simulations of plume particle dynamics and deposit spectra are rather idealized, estimates of the deposits' dimensions and their spectral trends are sufficient to identify the most promising observable signatures of recent activity on Europa. Below, we will consider the morphological indicators of plume deposits, and show that these are most likely to be detectable for very compact plumes whose resulting deposits accumulate rather quickly, and for plumes that are approximately 50 km tall, which are comprised of significantly more mass than plumes that are 10-25 km tall, and contain substantially more ice than the other large plumes we have considered. Next, we consider the spectral signatures of emplaced deposits and show that these may be a more promising approach for identifying fallout from larger and/or more transient plumes.

**4.1 Morphological Signatures of Plume Deposits**

Europa was last imaged by the Galileo spacecraft in 2002 [*Alexander et al.*, 2009], and the Europa Clipper spacecraft will reach the icy moon in the 2020s [*Phillips and Pappalardo*, 2014]. Assuming: (1) an approximately 25-year gap between the two missions, (2) that steady plume eruptions occur on approximately hour-long timescales [*Sparks et al.*, 2017], and (3) that deposits recognizable by cameras operating at visible wavelengths must be 1-10 m thick [*Quick et al.*, 2013], it is clear that deposits emplaced by very compact plumes will be easiest to identify on Europa (Table 4a-c). For example, with particle deposition rates near $9 \times 10^{-9}$ m/s , a 1 km tall plume could emplace a ~ 7 m thick deposit with 50% porosity, or a 34 m thick deposit with 90% porosity, if it erupted regularly, in the time between the two missions. Deposits emplaced by 1 km tall plumes would therefore be relatively thick and should be easily identifiable by high-resolution cameras, regardless of porosity. In general, we find that depending on deposit porosity, plumes that are less than 7 km tall would emplace deposits that could grow to be tens of meters thick in the time between the Galileo and Europa Clipper Missions. For example, applying the methodology outlined in Section 2.1, we find that an 800 m tall plume could emplace a 42 m thick deposit, assuming that the deposit has 90% pore space.

Conversely, our analysis shows that deposits emplaced by plumes that are ~10-25 km tall could be somewhat difficult for spacecraft cameras to detect. Owing to the relatively low mass of icy particles in these plumes (Table 3), and the large area over which deposits would be spread on the surface, deposition rates for these intermediate-sized plumes would be quite low, making it difficult for their deposits to accumulate sufficient mass in the time between the two missions to be detected. Even if detectable deposits were emplaced on the surface, they may be rendered unidentifiable by spacecraft cameras once they begin to coalesce and compress. For example, it would take between 38 and 98 years for compact deposits emplaced by plumes that are 10-25 km tall to grow to be 1 m thick (Table 4a). Further, although 10 km tall plumes could emplace ~ 3 m thick deposits in the time between the two missions, particles would only be able to form a layer this thick if the fraction of pore space between the particles remained very high. It is unknown whether icy particles in Europa's surface environment could resist compression for the 25 years between the two missions. As will be discussed in the next section, fresh surface deposits may be degraded by processes such as sintering and micrometeorite bombardment within a matter of decades [*Cooper et al.,*2001; *Tiscareno and Geissler,* 2003; *Carlson et al.* 2003; 2009].

In the case of plumes that are 50-300 km tall, we find that deposits would only accumulate enough mass to be detected by cameras operating at visible wavelengths for a very specific set of circumstances. Unless primarily composed of large particles, compact deposits (i.e., 50% pore space) produced by 50-200 km tall plumes would fail to reach 1m thickness in the time between the two missions. For example, deposits with 50% pore space that are emplaced by 50 km tall plumes may accumulate to ~ 2-4 m thick if primarily composed of particles that are 2-3$\mu$m in radius (Table 4b). Additionally, the $10^{-9}$-$10^{-8}$ m/s particle deposition rate for 50 km tall plumes would allow deposits with 90% pore space to be detectable to spacecraft cameras, regardless of the size of the constituent particles (Table 4b). Moreover, highly porous deposits that are primarily composed of larger particles could grow to be as much as 23 meters thick in the time between the two missions if emplaced by 50 km tall plumes (Table 4b). With the exception of plumes that are on the order of 1 km tall, deposits emplaced by 50 km plumes would have the highest probability of being detected by spacecraft cameras when compared to other large plumes (Table 4b). Indeed, even deposits that are primarily composed of particles with 0.5$\mu$m radii would accumulate enough mass to be detectable to visible imagers, provided that the deposits have a high percentage of pore space. This is not the case for plumes that are ≥100 km tall. Tables 4b-c show that high-porosity deposits emplaced by 100-300 km tall plumes would only be detectable if primarily composed of particles that are ≥ 1$\mu$m in radius, and that deposits emplaced by these plumes may only grow to be 10 m thick in the time between the two missions if they are composed of particles that are ≥ 3$\mu$m in radius.

If deposits must indeed be 1-10 m thick to be visible to spacecraft cameras, we find that that low-porosity deposits emplaced by 300 km tall plumes would not be detected by spacecraft cameras operating at visible wavelengths at all. Compact deposits emplaced by 300 km tall plumes would grow to be, at most, $10^{-3}$ mm thick in the intervening time between the Galileo and Europa Clipper missions. If these deposits consist of large particles, they would take at least 27 years to reach 1 m; if they primarily consist of small particles (e.g., 0.5 $\mu$m radii) it would take over 1800 years for them to emplace a 1 m thick layer on the surface (Fig. 4c). However, eruption of a 300 km tall plume could emplace a 1.4-5 m thick veneer of icy particles on Europa's surface if the deposit is highly porous, and is primarily composed of particles that are ≥ 2$\mu$m in radius (Table 4c).

Owing to their high ice to vapor ratios, and their small area of particle fallout, our calculations suggest that deposits emplaced by relatively small plumes, i.e., plumes that are < 7 km tall, would accumulate orders of magnitude faster and would be much thicker than their counterparts that are emplaced by larger plumes. Based on the likely morphology of deposits emplaced by large plumes, our analyses also suggest that it would be difficult for cameras operating at visible wavelengths to identify surface deposits that have been emplaced by plumes that are > 100 km tall unless the deposits are fresh, have been able to resist compression (i.e., remain highly porous), and/or are primarily composed of large particles. Features on Europa's surface brighten with age, and young features appear dark due to their larger grain sizes [*Geissler et al.*, 1998]. Hence assuming that they could accumulate enough mass, fresh deposits emplaced by large plumes might be identifiable as anomalous patches of dark deposits against an otherwise bright surface. Nevertheless, because these particles would be launched on trajectories that would carry them so far across the surface (Table 4b-c), it could be difficult to trace the resulting deposits back to their source locations. Further, as previously mentioned, any subsequent coalescence of deposits emplaced by large plumes could render them undetectable.

For all plume sizes considered, we find that deposits with 90% pore space accumulate 5 times faster and are therefore on average, 5 times thicker than their more compact counterparts that have only 50% pore space (Table 4). Accordingly, fresh, high-porosity deposits, and deposits that primarily consist of larger particles (i.e., $r_p$ = 2-3 $\mu$m) have the highest likelihood of being detected by spacecraft cameras, regardless of the size of the plumes that emplace them. This could suggest that if large plumes are common on Europa, and if these plumes leave behind surface deposits that are thick enough to be identified by spacecraft cameras, that their average particle sizes are at least on the order of 1$\mu$m, or that timescales for sintering or other processes that may facilitate their compression, are relatively long. Conversely, large plumes on Europa might consist of two particle populations, one which is composed of small particles that are entrained in the water vapor and may only be visible at wavelengths shorter than visible (e.g., UV), and another consisting of much larger particles that cluster near the vent. Enceladus' plumes are known to include particles with a range of sizes, with larger particles being more likely to be deposited on the moon's surface, while the smaller ones are more likely to escape into the E ring [*Porco et al.*, 2006; *Kempf et al.*, 2008; 2010; *Postberg et al.*, 2008; *Hedman et al.*, 2009; *Ingersoll and Ewald*, 2011]. The plume resulting from Io's Loki Patera also has two particle populations. Dust in the "outer plume" is $10^{-3}$-0.01 μm in radius and travels entrained in the $SO_2$ gas, and the dust in the "inner plume" is 1-1000 μm in radius, decouples from the gas, and clusters close to the vent [*Collins et al.*, 1981].

If Europan plumes consist of particle populations that are larger in size than those we considered, this will have an effect on our reported ice to vapor ratios. Ice to vapor ratios for the plumes we considered range from $I/V$ = 0.1 for 300 km tall plumes to $I/V$ = 332 for 1 km tall plumes (Table 3). Although the ice to vapor ratios for plumes that are likely to leave behind the most detectable deposits ($H$ < 10 km) are much greater than the $I/V$ estimated for plumes on Enceladus, they are commensurate with ice to vapor ratios for modeled eruptions on Europa and the Moon [*Wilson and Head*, 1983; *Fagents et al.*, 2000; *Quick et al.*, 2013]. In addition, ice to vapor ratios for 50 to 100 km tall Europan plumes are the same order of magnitude as dust to gas ratios for Io's Pele-type plumes and dust to gas ratios within the fine-grained dust component of Prometheus-type plumes (e.g., Thor and Loki) where the dust to gas ratio = 1 [*Geissler and McMillan*, 2008]. Further, ice to vapor ratios for 200-300 km tall plumes on Europa are within the range of plausible ice to vapor ratios reported for plumes on Enceladus [*Porco et al.*, 2006; *Kieffer et al.*, 2009; *Portyankina et al.*, 2017]. Moreover, Europa's plumes may differ significantly from Enceladus' in a variety of ways, including in intensity, output, and periodicity [*Rhoden et al.*, 2015]. It is therefore possible that Europa's plumes also differ from Enceladus' in terms of particle content. Bearing these details in mind, the ice to vapor ratios reported in Section 3, including the high ice to vapor ratios reported for small europan plumes, seem plausible.

We note however, that our analysis has assumed an idealized case where at least the smallest plume particles issue from the vent at velocities that are equal to the gas velocity. In reality, icy particles will issue from the vent at velocities that are somewhat slower than the gas, resulting in a low *I/V* at the foot of a plume. Additionally, as the gas and icy particles will reach different scale heights, *I/V* in the upper portions of plumes will be different from *I/V* at other locations. Moreover, observers may report a line-of-sight integrated *I/V*. For all of these reasons, the values of $n_{gas}$, $n_f$, $M_v$, $M$, and *I/V* listed for each plume in Table 3 may be substantially different from what is directly observed. This could also be the reason for the order of magnitude difference between the total water vapor mass calculated for the 200 km plume in Section 2 (Table 3), and the value reported (i.e., 1.46 x $10^6$ kg) in *Roth et al.* [2014]. Additionally, we have scaled total plume mass according to plume height, using water vapor column densities reported in *Sparks et al.* [2017] as a baseline. However, preliminary calculations using the methods described in Section 2 suggest that if the total mass of water vapor in all of Europa's plumes, regardless of their size, is on the order of $10^6$ kg, deposits that are composed of pure water ice and are emplaced by plumes ≥100 km tall, would not be easily detected. We note that although *Roth et al.* [2014] observed plumes that were 200 ± 100 km tall, and *Sparks et al.* [2016; 2017] reported repeat observations of a 50 km tall plume, both of these authors reported total water vapor masses of ~$10^6$ kg in the plumes they observed. This could indicate the total plume mass does not scale with plume height, or that the actual column densities of Europa's plumes differ from what is inferred from observation.

*Observational Constraints: Rhadamanthys and Androgeous Linea*

Previous workers have suggested that lineaments could be sites of recent geological activity on Europa [*Geissler et al.*, 1998]. Indeed, *Fagents et al.* [2000] and *Quick et al.* [2013]

considered that the low-albedo deposits which flank Rhadamanthys Linea (Fig. 11) and Androgeous Linea (Fig. 12) may be mantlings of cryoclastic particles that were emplaced by plumes. Rhadamanthys was imaged at 230 m/pixel and 1.6 km/pix during Galileo's E15 and G1 orbits, respectively, and Androgeous was imaged at 20 m/pix during Galileo's E6 orbit of Europa. The highest resolution images of Rhadamanthys, taken during Galileo's E15 orbit, provide the most accurate measurements of the dimensions of its low-albedo flanking deposits. The average radii of deposits flanking Rhadamanthys range from ~ 2 to 7 km, while the broadest portion of the deposit flanking Androgeous is approximately 3 km wide [*Fagents et al.*, 2000, Tables I & III]. Based on the visibility of preexisting topographic features beneath these deposits in Galileo imagery, and on Europa's $10^{-6}$ m/yr surface erosion rate [*Cooper et al.*, 2001], *Quick et al.* [2013] estimated these deposits to be 1-10 m thick. According to Table 4a, above, and Tables IIIa-IIIb of *Fagents et al.* [2000], the dimensions of these deposits are consistent with emplacement by plumes that were < 10 km tall.

Pre-existing topographic features are clearly visible beneath the low-albedo material flanking Rhadamanthys (Fig. 11). It is therefore reasonable to assume that the Rhadamanthys deposits are on the order of 1 m thick. According to Table 4a, the smallest Rhadamanthys deposits, i.e., those that are approximately 2 km in radius, could have been emplaced by 1 km tall plumes. Assuming 50% deposit porosity, we could expect each eruption to emplace a veneer of plume material with a maximum thickness of 0.12 mm. At these deposition rates, ~ 8333 eruptions would have to occur in order for a 1m thick deposit to accumulate. Assuming a deposit with 90% porosity, each eruption would emplace a veneer of plume material with a maximum thickness of 0.62 mm so that an approximately 1m thick veneer would accumulate after ~ 1613 eruptions. We find that it takes ~ 3.7 years for a 1 m thick deposit with 50% porosity to form, and ~ 9 months for a 1 m thick deposit with 90% porosity to form. In both cases, it would take a minimum of 6 eruptions per day to emplace just one of the smallest deposits in Fig. 11. At present, it is not clear if Europa's plumes are tidally modulated (e.g., see *Rhoden et al.* [2015]). Hence in all of these cases, the amount of time required to accumulate the modeled cryoclastic deposits is reported as a function of Earth days.

The largest Rhadamanthys deposit is ~ 7 km in radius [*Fagents et al.*, 2000]. Calculations using the procedures outlined in Section 2.1 suggest that its dimensions are consistent with having been emplaced by plumes that are 3.5 km tall. A 3.5 km tall plume could emplace a 0.04 mm thick deposit with 50% porosity each time it erupts. In this case, 25,000 eruptions would be required for a 1m thick deposit to form. We find that compact, 1m thick deposits emplaced by 3.5 km tall plumes would take approximately 13 years to accumulate. This suggests that 1923 eruptions per year, or 5 eruptions per day would be required to form the widest Rhadamanthys-flanking deposit (profile R18 in Table I of *Fagents et al* [2000]). Conversely, if the resulting deposit has 90% pore space, a 0.18 mm thick veneer would be emplaced on the surface after each eruption, and ~ 5556 eruptions would had to have occurred in order to for a 1m thick deposit to form. Calculations using the methods introduced in Section 2 suggest that 1m thick deposits could form in 2.6 years. This again suggests that over 2000 eruptions per year, or ~ 6 eruptions/day, would have been necessary to produce these deposits. Thus, regardless of porosity and the size of the resulting deposits, we find that a minimum of 5-6 eruptions per day would have been required to produce the Rhadamanthys deposits.

The deposits flanking Androgeous Linea (Fig. 12) appear to obscure more of the background plains than those that lie along Rhadamanthys. Because Androgeous was imaged at a much higher resolution than Rhadamanthys [*Fagents et al.*, 2000], it is possible that this heightened obscuration of the background plains is apparent rather than actual. Nevertheless, because of the decreased visibility of Androgeous' preexisting topographic features relative to Rhadamanthys' (Fig. 11), it appears that the dark deposits that flank the former are somewhat thicker than those flanking the latter. It is therefore possible that the low-albedo deposits lying along Androgeous Linea are ~10 m thick. Assuming plume eruptions took place at the edge of Androgeous, as noted in *Fagents et al.* [2000], we find that the 3 km wide portion of the Androgeous deposit could have been emplaced by plumes that were on the order of 0.8 km tall. If these deposits have 50% pore space, each eruption would emplace a veneer of material with a maximum thickness of approximately 0.16 mm, and ~ 63,000 eruptions would have to occur before a 10 m thick deposit could accumulate. We find that these eruptions would occur over, at most, 28 years' time, indicating that at a minimum, 186 eruptions per month, or just over 6 eruptions per day, would be required to form the broadest portion of the Androgeous deposit. Assuming deposits with 90% porosity, each eruption would emplace a veneer of plume material with a maximum thickness of 0.8 mm so that a 10 m thick veneer would accumulate after ~ 12,500 eruptions. In this instance, approximately 5.6 years would elapse before a 10 m thick deposit would accumulate on the surface. This equates to a minimum of 2232 eruptions per year, or once again, just over 6 eruptions per day, in order for these deposits to accumulate.

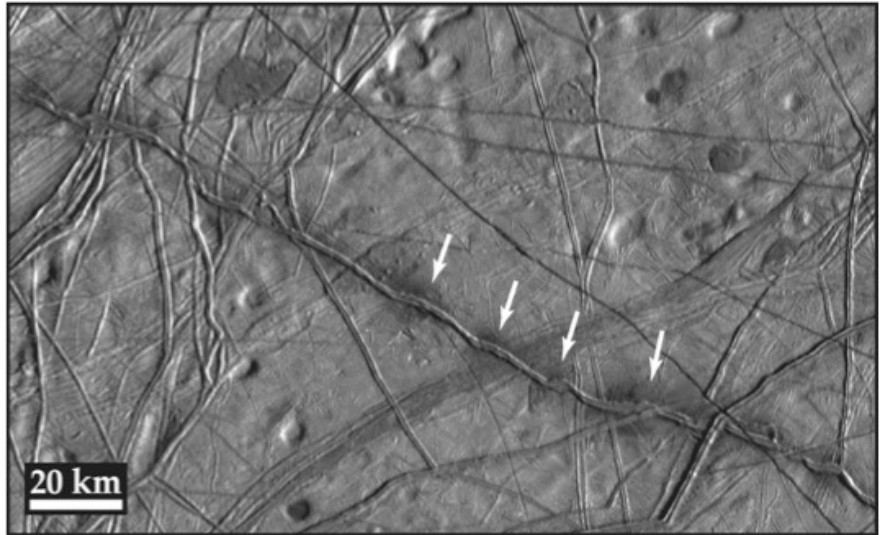

**Figure 11.** Rhadamanthys Linea from the Galileo spacecraft's E15 orbit of Europa at 230 m/pixel. The low-albedo deposits flanking this feature are indicated by the white arrows. These deposits range from approximately 2 – 7 km in radius (4 – 14 km in width) and may be cryoclastic mantlings that were emplaced by plumes.

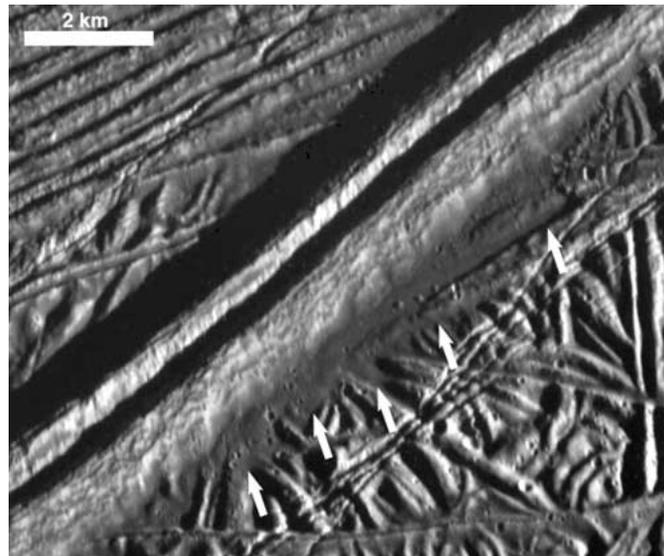

**Figure 12.** Segment of the prominent double ridge, Androgeous Linea. Image taken during Galileo's E6 orbit of Europa at 20 m/pixel. White arrows point to flanking low-albedo deposits that may be cryoclastic mantlings. According to *Fagents et al.* [2000], the broadest portion of this deposit is approximately 3 km wide.

### 4.2 Spectral and Photometric Signatures of Plume Deposits

While significant accumulation is needed to produce observable morphological signatures of plume deposits, spectral signatures can arise with much thinner deposits. In general, electromagnetic radiation only penetrates to depths a few times larger than the observed wavelength. Thus, a deposit that is only 10-100 $\mu$m thick would be enough to substantially affect the spectral properties of the surface at visible and near-infrared wavelengths. Even the 200-km tall plume considered above would produce a deposit of order 10 $\mu$m thick over the course of only a few days, so in principle even very short-lived eruptions could produce surface deposits with detectable trends in their spectral properties. These spectral signatures also extend over large regions and so do not require high spatial resolutions to detect, making them a promising way to search for deposits from large and/or intermittent plumes. Indeed, the spectral models discussed in Section 3.2 show that even if the plume particles do not have a distinctive composition, that the deposits can exhibit detectable spectral signatures. For example, the band depths of even a pure ice deposit from a 200-km high plume vary by a factor of two across a deposit hundreds of kilometers wide (see Figure 10). Detecting such a trend with a suitable near-infrared spectrometer would therefore only require spatial resolutions of order 10 km. More compact deposits would require correspondingly higher resolution, but the spectral variations associated with even a 10-km tall plume would still only require spatial resolutions on the order of 1 km.

Of course, in reality both the average size and the composition of plume particles can vary with distance from the vent, which could give rise to trends in wide variety of spectral and photometric parameters. For example, Europan plume deposits may contain significant fractions of non-ice material, perhaps in the form of condensed $CO_2$, $SO_2$, etc. [*Fagents et al.*, 2000], and the low-albedo deposits along Rhadamanthys and Androgeous may contain a significant fraction of salt species [*McCord et al.*, 1998; 1999; *Shirley et al.*, 2010]. Indeed, previous workers have suggested that low-albedo deposits on Europa's surface are likely composed of salts and other contaminants originating from the subsurface ocean or other internal liquid reservoirs [*Shirley et* *al.*, 2010; *Prockter et al.*, 2017]. If this is the case for the lineae-flanking deposits considered here, this would imply that the low-albedo spots along Rhadamanthys and Androgeous lineae are photometrically dark due to their composition. The density of NaCl is approximately equal to 2200 kg/m$^3$. Independent calculations using equations (2)-(5) with $\rho_p$ = 2200 kg/m$^3$ suggest that plumes ≥ 50 km tall could emplace deposits with radii on the order of 5-6 km, consistent with the dimensions of the largest Rhadamanthys deposits (see Table IIIa of *Fagents et al.*, 2000), if the erupted particles are primarily composed of salt, and are rather large, with average radii equal to 3$\mu$m. Eruptions of a 200 km tall plume could emplace a 15 $\mu$m thick deposit, with 50% pore space, composed primarily of 3$\mu$m salt particles, in about 7.5 years. A similar deposit with 90% pore space could form in just 1.5 years. Similar to the case for the emplacement of ice-dominated deposits, if the small Rhadamanthys deposits and the deposits flanking Androgeous Linea are primarily composed of salts, they were likely to have been emplaced by plumes < 4 km tall.

One potential challenge to identifying spectral and photometric signatures of plume deposits is that spectral signatures could be more transient than morphological signatures. Europa's surface spectra vary on a wide range of scales, with darker regions generally showing larger concentrations of non-icy materials (see *Carlson et al.*, 2009, and references therein). While some of these variations could be due to Europa's geological history and activity, others are almost certainly due to the surface being modified by the radiation environment. Radiation exposure can sinter grains, implant molecules, and sputter ice, while micrometeorites can mix surface deposits with underlying material. Various calculations indicate that these processes will mix or contaminate freshly exposed surfaces to depths of order 10-100 $\mu$m on timescales of order decades [*Cooper et al.,*2001; *Tiscareno and Geissler,* 2003; *Carlson et al.* 2003; 2009]. Hence the spectral and photometric signatures of surface deposits may become undetectable if they are more than a few decades old. Fortunately, the presence of dark patches around Rhadamanthys Linea, and the fact that dark terrains along lineae correspond to regions with higher fractions of non-ice material [*McCord et al.,* 1998; 1999], suggest that compact patches are emplaced regularly enough to maintain their distinct spectral and photometric properties. Alternatively, these deposits could have distinct compositional or structural features that are not easily erased by radiation exposure. If the former is correct, then this would imply that deposits from small plumes will likely be easier to detect both spectroscopically and morphologically. On the other hand, if the deposits have a persistent compositional signature, larger deposits could be easier to detect because they require lower resolution data. The possibility that 200 km tall plumes could emplace substantial deposits in the time between the Galileo and Europa Clipper missions, suggests that it is possible for plumes with ice to vapor ratios equal to 0.67 (Table 3), which is approximately equal to the maximum end of the *I/V* range *Ingersoll and Ewald* [2011] considered for plumes on Enceladus, to produce recognizable deposits on Europa's surface. For all other cases, the ice to vapor ratios of plumes that are likely to have produced the deposits are larger by several orders of magnitude.

## 5. Conclusions

In the absence of direct detection of plumes, plume deposits would provide the best evidence of recent geological activity on Europa and could also serve as important indicators for where to search for ongoing activity. We find that plumes that are less than 7 km tall are most likely to emplace deposits that are thick enough to be detected by spacecraft cameras operating at visible wavelengths. Mantlings emplaced by these plumes could accumulate to form deposits that are 1 - 10 m thick in as little as 7 months' time. If eruptive activity has occurred frequently on Europa since the Galileo Mission, these deposits could be substantial today, perhaps on the order of tens of meters thick. Moreover, we find that at most, ~ 5 - 6 eruptions/day of plumes that are 0.8-3.5 km tall are enough to produce the candidate cryoclastic mantlings flanking lineaments on Europa. Deposits emplaced by large plumes will be spread over large areas of the surface, but may accumulate enough mass to be detected by cameras operating at visible wavelengths if they are composed of particles > 0.5$\mu$m. Larger particles would cluster close to the source vent and the resulting deposits would be identifiable by cameras operating at visible wavelengths. Regardless of the size of the plumes that emplace them, we find that fresh, highly porous cryoclastic deposits and deposits that are primarily composed of particles with radii ≥ 2$\mu$m would be most visible from the point of view of spacecraft imagers, and hence most easily detected. Our analyses also indicate that any deposits that may have been emplaced by 100-300 km plumes would be visible to spacecraft cameras, provided that they are highly porous and/or composed of large particles. Nevertheless, within the parameter space explored here, we find that deposits emplaced plumes that are < 7 km tall, would be the easiest to detect.

Large candidate plumes that may be sporadic in nature have recently been observed on Europa. If these plumes are outliers, and most plumes on Europa are small in stature as suggested by previous modeling and image analysis, strategies for plume detection on the ocean moon should not only consider the potential periodicity of Europa's eruptions, but they should also consider the possibility that a significant number of Europa's plumes may be compact. Comprehensive plume search strategies should therefore include high-resolution imaging of low-albedo deposits that may have been emplaced via eruptive venting. These searches should pay special attention to the deposits that flank lineated features such as Rhadamanthys and Androgeous Lineae, and subsequent analyses should be carried out to determine if the albedos and/or dimensions of these deposits have changed in the time since the Galileo spacecraft first visited Europa.

The Europa Clipper spacecraft could constrain the amount of activity occurring along lineaments by acquiring high-resolution imagery of Androgeous and Rhadamanthys Linea. If the deposits that flank these features were indeed emplaced by plumes, then three scenarios are possible: (1) If Europa Clipper finds that the deposits have brightened and/or appear shrunken, then that would suggest that plume activity along these features ceased in the intervening decades between the two missions; (2) If imagery from Europa Clipper reveals that the albedo and dimensions of these deposits have remained unchanged since Galileo, then eruptions may have occurred along these lineaments in the time between the two missions, albeit at a steady state that allowed for the overall abundance of particles deposited onto the surface to remain constant; (3) If the deposits appear darker and wider in Europa Clipper imagery, then this might indicate that plume activity not only occurred continuously in the time between the two missions, but that plume output rates increased in the intervening years. This could take the form of material erupting from relatively small plumes for long periods of time between the two missions, or, could suggest that additional material was emplaced sporadically by large plumes in the intervening decades.

The Europa Imaging System (EIS) on Europa Clipper is well-suited to test the findings reported here and will be able to place improved constraints on the constitution of plume deposits. EIS will be able to detect surface color changes caused by the deposition of micron-sized plume particles, and stereo imaging by the Narrow Angle Camera (NAC) will be ideal for constraining the thicknesses of candidate plume deposits that flank lineaments and any other features [*Turtle et al.*, 2016; 2019]. Moreover, deposits emplaced by small plumes should be easily detected by EIS during its local- and regional-scale imaging campaigns. As previously mentioned, even if Europa Clipper does not directly detect plumes on Europa, their presence could be indirectly inferred via comparison of the dimensions of low-albedo deposits as they appear in Galileo imagery, with their appearance in high-resolution imagery acquired by EIS.

Meanwhile, the Mapping Imaging Spectrometer for Europa (MISE) will obtain near-infrared spectra of Europa's surface between 0.8 and 5.0 $\mu$m at 10 nm spectral resolution, and at spatial resolutions better than 10 kilometers on global scales, 500 meters at regional scales, and 25 meters at local scales [*Blaney et al.*, 2019]. This investigation will therefore be able to detect variations in both the surface composition and typical regolith particle sizes on scales comparable to many of the plume deposits considered here. Therefore, MISE will also place reasonable constraints on the levels of Europa's recent plume activity, particularly for large plumes whose morphological signatures may be difficult to discern.

**Acknowledgements**

We thank Sarah Fagents and an anonymous reviewer for feedback that improved the quality of this manuscript. We also thank Lionel Wilson for helpful discussions.
**References**
Alexander, C., Carlson, R., Consolmagno, G., Greeley, R., Morrison, D., 2009. The Exploration History of Europa. In: Pappalardo, R.T., McKinnon, W. B. and Khurana, K. (Eds.), Europa. University of Arizona Press, Tucson, pp. 3-26.
Batista, E. R., Ayotte, P., Bilic, A., Kay, B. D., Jonsson, H., 2005. What Determines the Sticking Probability of Water Molecules on Ice? Physical Review Letters 95, 223201.
Berg, J. J., Goldstein, D. B., Varghese, P. L., Trafton, L. M., 2016. DSMC Simulation of Europa Water Vapor Plumes. Icarus 277, 370-380.
D.L. Blaney, C. Hibbitts, R.O. Green, R.N. Clark, J.B. Dalton, A.G. Davies, Y. Langevin, J.I. Lunine, M.M. Hedman, T.B. McCord, S.L. Murchie, C. Paranicas, F.P. Seelos, J.M. Soderblom, S. Diniega, M. Cable, D. Thompson, C. Bruce, A. Santo, R. Redick, D. Hahn, H. Benderm, B. Van Gorp, J. Rodriguez, P. Sullivan, T. Nevillem S. Lundeen, M. Bowers, K. Ryan, J. Hayes, B. Bryce, R. Hourani, E. Zarate, L.B. Moore, K. Maynard, I.M. McKinley, D. Johnson, P. Aubuchon, J. Fedosi, R. Wehbe, R. Calvet, P. Mouroulis, V. White, D. Wilson. 2019. The Europa Clipper Mapping Imaging Spectrometer for Europa (MISE): Using Compositional Mapping to Understand Europa. 50th Lunar and Planetary Science Conference, Abstract #2132.

Bramson, A. M., Phillips, C. B., Emery, J. C., 2011. A Search for Geologic Activity on Jupiter's Satellites. 42nd Lunar and Planetary Science Conference, Abstract #1606.
Carlson, R.W., Calvin, W. M., Dalton, J.B., Hansen, G. B., Hudson, R.L., Johnson, R. E., McCord, T. B., Moore, M. H., 2009. Europa's Surface Composition. In: Pappalardo, R. T., McKinnon, W. B. and Khurana, K. (Eds.),Europa. University of Arizona Press, Tucson, pp. 283-327.
Carlson, R.W, Anderson, M. S., Johnson, R. E., Schulman, M. B., A.H. Yavrouian, A. H., 2002. Sulfuric Acid production on Europa: The Radiolysis of Sulfur in Water Ice. Icarus, 157, 456-463.
Collins, S. A., 1981. Spatial Color Variations in the Volcanic Plume at Loki, on Io. Journal of Geophysical Research 86, 8621-8626.
Cook, A.F., Danielson, G.E., Jewitt, D.C., Owen, T., 1981.Dust Observations in the Jovian System. Advances in Space Research 1, COSPAR, Great Britain, 99–101.
Cooper, J.F., Johnson, R. E., Mauk, B. H., Garrett, H.B., Gehrels, N., 2001. Energetic Ion and Electron Irradiation of the Icy Galilean Satellites. Icarus 149, 133-159.
Crawford, G. D., Stevenson, D. J., 1988. Gas-Driven Water Volcanism and the Resurfacing of Europa. Icarus 73, 66-79.
Cuffey, K.M., Paterson, W. S. B., 2010. The Physics of Glaciers, Fourth Edition, Elsevier, USA.
Cuzzi, J. N., Estrada, P. R., 1998. Compositional Evolution of Saturn's Rings Due to Meteoroid Bombardment. Icarus 132, 1–35.
Degruyter, W., Manga, M., 2011. Cryoclastic Origin of Particles on the Surface of Enceladus. Geophysical Research Letters 38, L16201.
Europa Study Team, 2012. Europa Study 2012 Report. Task Order NMO11062, NASA/Jet Propulsion Laboratory.
Fagents, S.A., Greeley, R., Sullivan, R.J., Pappalardo, R.T., Prockter, L.M., 2000. Cryomagmatic Mechanisms for the Formation of Rhadamanthys Linea, Triple Band Margins, and Other Low Albedo Features on Europa. Icarus 144, 54-88.
Fagents, S. A., 2003. Considerations for Effusive Cryovolcanism on Europa: The Post-Galileo Perspective. Journal of Geophysical Research 108, 5139.
Gaidos, E. J., Nimmo, F., 2000. Tectonics and Water on Europa. Nature 405, 637.
Johnson, T. V., Matson, D. L., Blaney, D. L., Veeder, G. J., 1995. Stealth Plumes on Io. Geophysical Research Letters 22, 3293-3296.
Geissler, P.E., Greenberg, R., Hoppa, G., McEwen, A., Tufts, R., Phillips, C., Clark, B., Ockert-Bell, M., Helfenstein, P., Burns, J., Veverka, J., Sullivan, R., Greeley, R., Pappalardo, R.T., Head, J.W., Belton, M.J.S., Denk, T., 1998. Evolution of Lineaments of Europa: Clues from Galileo Multispectral Imaging Observations. Icarus 135, 107-126.
Geissler, P., McEwen, A., Phillips, C., Keszthelyi, L., Spencer, J., 2004. Surface Changes on Io During the Galileo Mission. Icarus 169, 29-64.
Geissler, P. E., Goldstein, D. B., 2006; Plumes and their Deposits. In: Lopes, R. M. C., and Spencer, J. R. (Eds.), Io After Galileo: A New View of Jupiter's Volcanic Moon. Springer Praxis Books, New York, NY, pp. 163-192.
Geissler, P. E., McMillan, M. T., 2008. Galileo Observations of Volcanic Plumes on Io. Icarus 197, 505-518.
Glaze, L. S., Baloga, S. M., 2000. Stochastic-Ballistic Eruption Plumes on Io. Journal of Geophysical Research 105, 17579-17588.
Hapke, B. 1981. Bidirectional Reflectance Spectroscopy. I - Theory. Journal of Geophysical

Research 86, 3039–3054.

Hapke, B. 1993. Theory of Reflectance and Emittance Spectroscopy. Cambridge University Press, Cambridge, UK.

Hedman, M. M., Nicholson, P. D., Showalter, M. R., Brown, R. H., Buratti, B. J., Clark, R. N., 2009. Spectral Observations of the Enceladus Plume with Cassini-VIMS. The Astrophysical Journal 693, 1749–1762.

Hedman, M.M., Nicholson, P. D., Cuzzi, J. N., Clark, R. N., Filacchione, G., Capaccioni, F., Ciarniello, M., 2013. Connections Between Spectra and Structure in Saturn's Main Rings Based on Cassini VIMS Data. Icarus 223, 105-130.

Hedman, M. M., Dhingra, D., Nicholson, P. D., Hansen, C. J., Portyankina, G., Ye, S., Dong, Y., 2018. Spatial Variations in the Dust-to-Gas Ratio of Enceladus' Plume. Icarus 305, 123-138.

Hsu, H.-W., Postberg, F., Sekine, Y., Shibuya, T., Kempf, S., Horányi, M., Juhász, A., Altobelli, N., Suzuki, K., Masaki, Y., Kuwatani, T., Tachibana, S., Sirono, S.-i., Moragas-Klostermeyer, G., Srama, R., 2015. Ongoing Hydrothermal Activities within Enceladus. Nature 519, 207-210.

Ingersoll, A. P., Ewald, S. P., 2011. Total Particulate Mass in Enceladus Plumes and Mass of Saturn′s E ring Inferred from Cassini ISS Images. Icarus 216, 492–506.

Jia, X., Kivelson, M. G., Khurana, K. K., Kurth, W. S., 2018. Evidence of a Plume on Europa from Galileo Magnetic and Plasma Wave Signatures. Nature Astronomy 2, 459-464.

Kempf, S., Beckmann, U., Moragas-Klostermeyer, G., Postberg, F., Srama, R., Economou, T., Schmidt, J., Spahn, F., Grun, E., 2008. The E-ring in the Vicinity of Enceladus: I. Spatial Distribution and Properties of Ring Particles. Icarus 193, 420–437.

Kempf, S., Beckmann, U., Schmidt, J., 2010. How the Enceladus Dust Plume Feeds Saturn's E Ring. Icarus 206, 446-457.

Kieffer, S.W., Lu, X., McFarquhar, G., Wohletz, K.H., 2009. A Redetermination of the Ice/Vapor Ratio of Enceladus′ Plumes: Implications for Sublimation and the Lack of a Liquid Water Reservoir. Icarus 203, 238–241.

Mahieux, A., Goldstein, D. B., Varghese, P. L., Trafton, L. M., 2019. Parametric Study of Water Vapor and Water Ice Particle Plumes Based on DSMC Calculations: Application to the Enceladus Geysers. Icarus 319, 729-744.

Mastrapa, R. M., Sandford, S. A., Roush, T. L., Cruikshank, D. P., Ore, C.M.D., 2009. Optical Constants of Amorphous and Crystalline $H_2O$-ice: 2.5-22 $\mu$ m (4000-455 $cm^{-1}$) Optical Constants of $H_2O$-ice. The Astrophysical Journal 701, 1347-1356.

McCord, T.B., Hansen, G. B., Fanale, F.P., Carlson, R.W., Matson, D. L., Johnson, T. V., Smythe, W. D., Crowley, J. K., Martin, P. D., Ocampo, A., Hibbitts, C.A., Granahan, J. C., 1998. Salts on Europa's Surface Detected by Galileo's Near-Infrared Mapping Spectrometer. Science 280, 1242.

McCord, T.B., Hansen, G. B., Matson, D. L., Johnson, T. V., Crowley, J. K., Fanale, F. P., Carlson, R.W., Smythe, E.D., Martin, P.D., Hibbitts, C. A., Granahan, J. C., Ocampo, A. Hydrated Salt Minerals on Europa's Surface from the Galileo Near-Infrared Mapping Spectrometer (NIMS) investigation. Journal of Geophysical Research 104, 11827-11852.

Nordheim, T. A., Hand, K. P., Paranicas, C., 2018. Preservation of Potential Biosignatures in the Shallow Subsurface of Europa. Nature Astronomy 2 673-679.

Pappalardo, R., Senske, D., Prockter, L., Paczkowski, B., Vance, S., Rhoden, A., Goldstein, B., Magner, T., Cooke, P., 2015. Science Objectives for the Europa Clipper Mission Concept:

Investigating the Potential Habitability of Europa. European Planetary Science Conference 10, Abstract #EPSC2015-156.
Parfitt, E. A., Wilson, L., 2008. Fundamentals of Physical Volcanology. Blackwell Publishing, Malden, MA.
Phillips, C. B., McEwen, A.S., Hoppa, G.V., Fagents, S.A., Greeley, R., Klemaszewski, J.E., Pappalardo, R.T., Klaasen, K.P., Breneman, H.H., 2000. The Search for Current Geologic Activity on Europa. Journal of Geophysical Research 105, 22579-22597.
Phillips, C. B., Pappalardo, R. T., 2014. Europa Clipper Mission Concept: Exploring Jupiter's Ocean Moon. Eos 95, 165-167.
Porco, C.C., Helfenstein, P., Thomas, P. C., Ingersoll, A.P., Wisdom, J., West, R., Neukum, G., Denk, T., Wagner, R., Roatsch, T., Kieffer, S., Turtle, E., McEwen, A., Johnson, T.V., Rathbun, J., Veverka, J., Wilson, D., Perry, J., Spitale, J., Brahic, A., Burns, J.A., DelGenio, A.D., Dones, L., Murray, C.D., Squyres, S., 2006. Cassini Observes the Active South Pole of Enceladus. Science 311,1393–1401.
Porco, C. C., Dones, L., Mitchell, C., 2017. Could it Be Snowing Microbes on Enceladus? Assessing Conditions in its Plume and Implications for Future Missions. Astrobiology 17, 876-901.
Portyankina, G., Hedman, M. M., Hansen, C. J., Esposito, L. W., Aye, K. -M., Dhingra, D., 2017. Simultaneous Cassini UVIS and VIMS Solar Occultation Observations: Modeling Insights. 48th Lunar and Planetary Science Conference. Abstract #2418.
Postberg, F., Kempf, S., Hillier, J.K., Srama, R., Green, S.F., McBride, N., Grun, E., 2008. The E-Ring in the Vicinity of Enceladus: II. Probing the Moon′s Interior - The Composition of E-Ring Particles. Icarus 193, pp.438–454.
Postberg, F., Schmidt, J., Hillier, J., Kempf, S., Srama, R., 2011. A Salt-Water Reservoir as the Source of a Compositionally Stratified Plume on Enceladus. Nature 474, 620-622.
Postberg, F., Khawaja, N., Abel, B., Choblet, G., Glein, C. R., Gudipati, M., Henderson, B. L., Hsu, H.-.W., Kempf, S., Klenner, F., Moragas-Klostermeyer, G., Magee, B., Nölle, L., Perry, M., Reviol, R., Schmidt, J., Srama, R., Stolz, F., Tobie, G., Trieloff, M., Waite, J. H., 2018. Macromolecular Organic Compounds from the Depths of Enceladus. Nature 558, 564-567.
Poulet, F., Cuzzi, J. N., Cruikshank, D. P., Roush, T., Dalle Ore, C.M, 2002. Comparison Between the Shkuratov and Hapke Scattering Theories for Solid Planetary Surfaces: Application to the Surface Composition of Two Centaurs. Icarus 160, 313–324.
Prockter, L. M., Shirley, J. H., Dalton, J. B. III, Kamp, L., 2017. Surface Composition of Pull-Apart Bands in Argadnel Regio, Europa: Evidence of Localized Cryovolcanic Resurfacing During Basin Formation. Icarus 285, 27-42.
Quick, L. C., Barnouin, O. S., Prockter, L. M., Patterson, W. G., 2010. Constraints on the Detection of Cryovolcanic Plumes on Europa. 41st Lunar and Planetary Science Conference, Abstract #2247.
Quick, L. C., Barnouin, O. S., Prockter, L. M., Patterson, W. G., 2013. Constraints on the Detection of Cryovolcanic Plumes on Europa. Planetary and Space Science 86, 1-9.
Rathbun, J., Spencer, J., 2018. A Closer Look at Galileo Thermal Data from Possible Plume Sources Near Pwyll, Europa. 50th Meeting of the AAS Division for Planetary Sciences, Abstract #403.06
Rathbun, J., Spencer, J., 2019. Proposed Plume Source Regions on Europa: No Evidence for Endogenic Thermal Emission. Icarus, In Press.

Rhoden, A. R., Hurford, T. A., Roth, L., Retherford, K., 2015. Linking Europa's Plume Activity to Tides, Tectonics, and Liquid Water. Icarus 253, 169-178.
Roth, L., Saur, J., Retherford, K.D., Strobel, D.F., Feldman, P.D., McGrath, M.A., Nimmo, F., 2014. Transient Water Vapor at Europa's South Pole. Science 343, 171-174.
Schmidt, J., Brilliantov, N., Spahn, F., Kempf, S., 2008. Slow Dust in Enceladus′ Plume from Condensation and Wall Collisions in Tiger Stripe Fractures. Nature 451, 685–688.
Schmidt, B. E., Blankenship, D. D., Patterson, G. W., Schenk, P. M., 2011. Active Formation of Chaos Terrain Over Shallow Subsurface Water on Europa. Nature 479, 502-505.
Shaw, R. A., Lamb, D., 1999. Experimental Determination of the Thermal Accommodation and Condensation Coefficients of Water. Journal of Chemical Physics 111, 10659-10663.
Shirley, J. H., Dalton, J. B. III, Prockter, L. M., Kamp, L. W., 2010. Europa's Rigid Plains and Smooth Low Albedo Plains: Distinctive Compositions and Compositional Gradients at the Leading Side-Trailing Side Boundary. Icarus 210, 358-384.
Shkuratov, Y., Starukhina, L, Hoffmann, H., Arnold, G., 1999. A Model of Spectral Albedo of Particulate Surfaces: Implications for Optical Properties of the Moon. Icarus 137, 235–246.
Spahn, F., Schmidt, J., Albers, N., Horning, M., Makuch, M., Seib, M., Kempf, S., Srama, R., Dikarev, V., Helfert, S., Moragas-Klostermeyer, G., Krivov, A.V., Sremcevic, M., Tuzzolino, A.J., Economou, E., Grun, E., 2006. Cassini Dust Measurements at Enceladus and Implications for the Origin of the E Ring. Science 311, 1416–1418.
Southworth, B. S., Kempf, S., Schmidt, J., 2015. Modeling Europa's Dust Plumes. Geophysical Research Letters 42, 10451-10458.
Sparks, W. B., Hand, K. P., McGrath, M. A., Bergeron, E., Cracraft, M., Deustua, S. E., 2016. Probing for Evidence of Plumes on Europa with HST/STIS. The Astrophysical Journal 829, 121.
Sparks, W. B., Schmidt, B. E., McGrath, M. A., Hand, K. P., Spencer, J. R., Cracraft, M., Deustua, S. E., 2017. Active Cryovolcanism on Europa? The Astrophysical Journal Letters 839, L18.
Strom, R.G., Schneider, N.M., Terrile, R.J., Cook, A. F., Hansen, C., 1981.Volcanic Eruptions on Io. Journal of Geophysical Research 86, 8593–8620.
Tiscareno, M.S., Geissler, P. E., 2003. Can Redistribution of Material by Sputtering Explain the Hemispheric Dichotomy of Europa? Icarus 161, 90-101.
Turtle, E.P., McEwen, A.S., Barr, A. C., Collins, G.C., Fletcher, L. N., Hansen, C, J., Hayes, A., Hurford, T.A., Kirk, R.L., Nimmo, F., Patterson, G.W., Quick, L.C., Soderblom, J.M., Thomas, N., Ernst, C., 2016. The Europa Imaging System (EIS): High-Resolution Imaging and Topography to Investigate Europa's Geology, Ice Shell, and Potential for Current Activity. 47th Lunar and Planetary Science Conference, Abstract #1626.
Turtle, E. P., McEwen, A. S., Collins, G. C., Dauber, I. J., Ernst, C. M., Fletcher, L., Hansen, C. J., Hawkins, S. E., Hayes, A. G., Humm, D., Hurford, T. A., Kirk, R. L., Kutsop, N., Barr Mlinar, A. C., Nimmo, F., Patterson, G. W., Phillips, C. B., Pommerol, A., Prockter, L., Quick, L. C., Reynolds, E. L., Slack, K. A., Soderblom, J. M., Sutton, S., Thomas, N., Bland, M., 2019. The Europa Imaging System (EIS): High-Resolution, 3-D Insight into Europa's Geology, Ice Shell, and Potential for Current Activity. 50th Lunar and Planetary Science Conference, Abstract #3065.
Wilson, L., Head, J. W. III, 1983. A Comparison of Volcanic Eruption Processes on Earth, Mars, Moon, Io, and Venus. Nature 302, 663-669.
Ye, S.-Y., Gurnett, D.A., Kurth, W.S., Averkamp, T.F., Kempf, S., Hsu, H.-W., Srama, R., Grun,

E., 2014. Properties of Dust Particles near Saturn Inferred from Voltage Pulses Induced by Dust Impacts on Cassini Spacecraft. Journal of Geophysical Research: Space Physics 119, 6294-6312.

Yeoh, S. K., Chapman, T. A., Goldstein, D. B., Varghese, P. L., Trafton, L. A., 2015. On Understanding the Physics of Enceladus South Polar Plume Via Numerical Simulation. Icarus 253, 205-222.

Zhang, J., Goldstein, D. B., Varghese, P. L., Trafton, L., Moore, C., Miki, K., 2004. Numerical Modeling of Ionian Volcanic Plumes with Entrained Particulates. Icarus 172, 479-502.